\documentclass[aps,pre,twocolumn,floatfix,superscriptaddress]{revtex4-1}

\usepackage{graphicx}

\usepackage{amsfonts,amsmath}
\usepackage{appendix}
  \usepackage{psfig}

\begin{document}

\title{Pair approximation for the noisy threshold $q$-voter model}

\author{A. R. Vieira$^1$, Antonio F. Peralta$^2$, Raul Toral$^2$, Maxi San Miguel$^2$, C. Anteneodo$^{3}$} 
\address{$^1$Department of Physics, PUC-Rio, Rua Marqu\^es de S\~ao Vicente, 225, 22451-900, Rio de Janeiro, Brazil\\
$^2$IFISC, Instituto de F{\'\i}sica Interdisciplinar y Sistemas Complejos (CSIC-UIB), Campus Universitat de les Illes Balears, 07122 Palma de Mallorca, Spain\\
$^3$National Institute of Science and Technology for Complex Systems (INCT-SC), Rio de Janeiro, Brazil}

 %

\begin{abstract}
In the standard $q$-voter model, a given agent can change its opinion only if 
there is a full consensus of the opposite opinion within 
a group of influence of size $q$. 
A more realistic extension is the threshold $q$-voter, where a minimal agreement 
(at least $0<q_0\le q$ opposite opinions) is sufficient to flip the central agent's opinion, 
including also the possibility of independent (non conformist) choices. 
Variants of this model including non-conformist behavior have been previously 
studied in fully connected networks (mean-field limit). 
Here we investigate its dynamics in random networks. 
Particularly, while in the mean-field case it is irrelevant whether 
repetitions in the influence group are allowed, we show that this is not the case in networks, and we study the impact of both cases, with or without repetition. 
Furthermore, the results of computer simulations are compared with the predictions 
of the pair approximation derived for uncorrelated networks of arbitrary degree distributions. 
\end{abstract}

\maketitle

\section{Introduction}

The voter model is a paradigmatic exponent of opinion dynamics in a binary scenario, 
where individuals have to choose between two alternatives, e.g., for or against. 
According to this model, a randomly chosen agent can flip its opinion by imitation, 
copying the choice of one of its contacts~\cite{review,voter1,voter2}. 

An enriched variant of the voter model is the $q$-voter 
proposed by Castellano et al.~\cite{q-voter}, where an agent is persuaded 
not by one but by $q$ neighbors in the network of contacts. 
However, the consensus among $q$ contacts required for 
changing opinion seems too restrictive. 
In a real scenario, a simple majority or even a minimal number may be enough. 
This can be expressed by the introduction of 
a threshold~\cite{threshold-sznajd,threshold-celia-allan,sznajd-plos}, 
which allows a flip of opinion when there is 
a minimal number of $q_0$ contacts sharing the opposite opinion 
within the influence group of size $q$.
 
It is also realistic to include the possibility of stochastic behavior 
given by anti-conformist or nonconformist trends~\cite{q-voter-noises,q-voter-zealots}. In the original version of the $q$-voter model~\cite{q-voter}, 
a parameter $\varepsilon$ is introduced which has a hybrid interpretation 
in terms of nonconformism. 
But alternative rules have been proposed~\cite{sznajd-plos}. 
For instance, the chosen agent can flip its opinion state ignoring the neighbors, 
with a given probability $p$, 
while the conformist response following the threshold dynamics occurs with probability $1-p$.
This is the variant of the threshold $q$-voter that we consider in this paper.

Another relevant ingredient in opinion dynamics is the structure of 
the network of contacts. 
However, few works on $q$-voter models go beyond the mean-field dynamics. 
Among them, we find the standard $q$-voter in complex 
networks~\cite{qvoterRRN,qvoterPA} and in multiplex networks~\cite{qvoter_multiplex,qvoter_multiplexPA}, 
as well as the non-linear noisy voter~\cite{antonioChaos,antonioNJP}, 
related to the $q$-voter, for which analytical results 
were obtained through the pair approximation (PA). 
In order to investigate the threshold $q$-voter on random networks we use the PA approximation 
complemented by Monte Carlo simulations.

Furthermore, 
a feature that is not significant in the mean-field (or globally coupled) case 
but relevant when we introduce network structure 
is the possibility of repetition or not when choosing $q$ agents amongst $k$ neighbors. 
Therefore, we also investigate the effects of repetitions on the critical 
portrait of this voter model. 

In Sec.~\ref{sec:model} we describe the opinion dynamics, and revisit the mean-field 
limit in Sec.~\ref{sec:MF}. 
We present analytical results using the PA in Sec.~\ref{sec:PA}, 
and comparison with simulations in Sec.~\ref{sec:results}. 
Finally, concluding remarks are addressed in Sec.~\ref{sec:remarks}. Details of the calculations can be found in Appendix A and Appendix B.

\section{Model}
\label{sec:model}
Let us consider that individuals in a network of size $N$ can hold any of two opinions ($\oplus$ and $\ominus$), $n$ of them with opinion $\oplus$ and $N-n$ the opposite one $\ominus$.
The state of each individual can change according to the following 
algorithm:
We randomly select an individual (node) $i$ in the network. With probability $1-p$ this opinion is subject to change in conformity with an 
influence group, while with probability $p$ it acts independently of its neighbors, flipping its state with chance 1/2. 
In the conformity rule, we select $q$ neighbors of the central node $i$. 
If in this set there are at least $q_0$ elements that share the opposite opinion of node $i$, then its opinion state is flipped. The particular case $q_0=q$ represents the $q$-voter model, where unanimity is required in order to influence opinions \cite{q-voter}. The still more particular case $q_0=q=1$ corresponds to the standard well-known ``noisy-voter model'' (Kirman model) \cite{noisy-voter1,noisy-voter2}. The noise-parameter $p$ is a measure of the degree of independence of a node with respect to its neighbors, and we name this particular set of rules as the ``noisy threshold $q$-voter model''.

Let us remark that in the selection of the influence set, the $q$ nodes can be either distinct (i.e. chosen without repetitions) or one can allow for repetitions, which reflects interactions that are more frequent or intense. As we will see, the existence or not of repetitions in the selection of the influence set is a relevant ingredient of the model.

We note also that, at variance with the version discussed in~\cite{threshold-celia-allan}, 
where a different type of noise was considered~\cite{q-voter}, the systems discussed here do not show the existence of absorbing states ($n=0$ and $n=N$) where the systems get trapped.

In the following sections, analytical results are obtained for networks of arbitrary degree distribution. 
Simulations are carried out mainly in random-regular networks~\cite{survey-rrn,algorithm-rrn} 
but, Erd\H{o}s-R\'enyi graphs~\cite{Erdos1959a,Barabasi2002a} 
and networks with a power-law degree distribution~\cite{config1,config2} are also considered.

\section{Mean-field solution}
\label{sec:MF}

In this Section we present the mean-field (MF) scenario, with the idea of defining the quantities needed to develop the more sophisticated pair approximation scheme in Sec.~\ref{sec:PA} and to provide a framework of comparison for our analytical and numerical results in structured networks.

\begin{figure}[h!]
\begin{center} 
\includegraphics[scale=0.26,angle=-90]{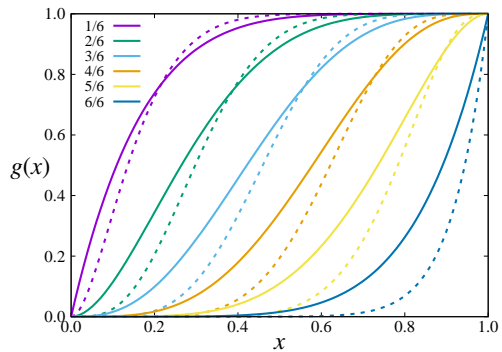} \\
\end{center}
\caption{ Flipping probability $g_1(x;q,q_0)$, in the absence of noise ($p=0$),
 for values of $q_0/q$ indicated in the legend, with $q=6$ (solid lines) and $q=12$ (dashed lines). 
}
\label{fig:gx}
\end{figure}

Let us then consider a fully connected network of size $N$. The stochastic evolution of the collective (or average) opinion can be described 
by knowing the transition rates $w(n\to n^\prime)$ between collective states characterized, respectively, by $n$ and $n^\prime$ agents with opinion $\oplus$. The (non-null) transition rates are
\begin{eqnarray}\label{eq: nonnull tax}
w(n\to n-1) & = & n \, G(1-x) , \\ \nonumber
w(n\to n+1) & = & (N-n) \, G(x) , 
\end{eqnarray}
where $x \equiv n/N$  
and $ G(x)$ [resp. $G(1-x)$] is the conditional probability that a sorted agent with opinion $\ominus$ [resp. $\oplus$] flips its opinion. 
In the extension of the $q$-voter dynamics with threshold $q_0$ and independence level $p$, 
this flipping probability is given by the governing rules specified in Section \ref{sec:model}, see for instance, reference \cite{sznajd-plos},
\begin{equation}
 \label{eq:G}
G(x;q,q_0,p)=(1 - p)g_1(x;q,q_0) + p/2 \, ,
\end{equation}
where 
\begin{eqnarray} \label{eq:gx}
&&g_1(x;q,q_0) = \sum_{j = q_0}^{q} {{q}\choose{j}} \, x^j (1-x)^{q-j} .
\end{eqnarray}
is the cumulative probability of sorting at least $q_0$ agents with opposite opinion amongst $q$ neighbors. As explained in Appendix A, it turns out that $g_1(x;q,q_0)$ can be written in terms of hypergeometric functions.
Eq. (\ref{eq:gx}) refers to the case where one allows repetitions in the selection of the $q$ neighbors. However, in the fully-connected network with $N\gg q$, the probability of repetition is very small and hence in the thermodynamic limit $N\to\infty$ the results coincide for both selection options. Function $g_1(x;q,q_0)$ is depicted in Fig.~\ref{fig:gx} for several values of $q$ and $q_0$.

The time evolution of the average value $\langle x \rangle$ can be obtained from the rate equation
\begin{equation}
 \label{eq:drift}
\frac{dx}{dt} = (1-x)G(x) - x G(1-x) \equiv \upsilon(x;p,q,q_0) \, ,
\end{equation}
where we have used the deterministic approximation 
$\langle x^{m} \rangle = \langle x \rangle^{m}$ 
(which is of general validity in the thermodynamic limit $N \rightarrow \infty$ \cite{peralta-moments}) 
and short notation $\langle x \rangle \equiv x$. Note that, due to the symmetries of the model, the function $\upsilon(x)$ can be expanded in a Taylor series containing only odd powers of $(x-1/2)$. Furthermore, the steady-state probability of finding $n$ agents in state $\oplus$ can be written, in the limit of large $N$, in a large-deviation form $P_\textrm{st}(n)\propto e^{-NV(n/N)}$, with a potential function~\cite{peralta:JSTAT}
\begin{equation}
V(x)=\int^xdx\log\left[\frac{xG(1-x)}{(1-x)G(x)}\right].
\end{equation}
The fixed points of Eq.~(\ref{eq:drift}), satisfying $\upsilon(x;p) =0$ coincide with the extrema of the potential $V(x)$, such that stable (resp. unstable) fixed points are minima (resp. maxima) of $V(x)$. In the case of coexistence of multiple stable fixed points, the macroscopically observed configurations (phases) correspond to the absolute minimum of the potential, while relative minima are metastable phases. The fixed points are obtained as the roots $x(p)$ of the equation 
\begin{equation}
 \label{eq:extreme}
 p = \frac{x {g_1}(1-x) - (1-x) {g_1}(x)}{x {g_1}(1-x) - (1-x) {g_1}(x) + (1/2-x)} \, .
\end{equation}
We now discuss the different solutions and their stability. 

The trivial solution $x(p)=1/2$, that corresponds to a disordered (D) phase, exists for all parameter values and it is stable for $p>p_c$ and unstable for $p<p_c$. The stability boundary can be obtained by taking the limit $x\to 1/2$ in Eq.~(\ref{eq:extreme}), yielding
\begin{equation} \label{pc_MF}
p_c^{-1} = 1 + \dfrac{2^{q-1} \Gamma(q_0+1)\Gamma(q - q_0 +1) }
{\Gamma(q+1) [q_0- {_2F_1}(1, q_0-q, q_0+1, -1)]} ,
\end{equation}
This equation generalizes the mean-field result for the $q$-voter model~\cite{qvoterPA}, $p_c^{-1} = 1+ \frac{2^{q-1}}{q-1}$, recovered by setting $q_0=q$ in Eq.~(\ref{pc_MF}). We note that $p_c$ becomes equal to zero for $q_0\le q_0^*$ where $q_0^*$ is the value that cancels the denominator of the right-hand-side of Eq.~(\ref{pc_MF}), namely $q_0^*={_2F_1}(1, q_0^*-q, q_0^*+1, -1)$. This equation yielding the value $q_0^*(q)$ has to be solved numerically.

\begin{figure}[h!]
\begin{center} 
\includegraphics[scale=0.46,angle=0]{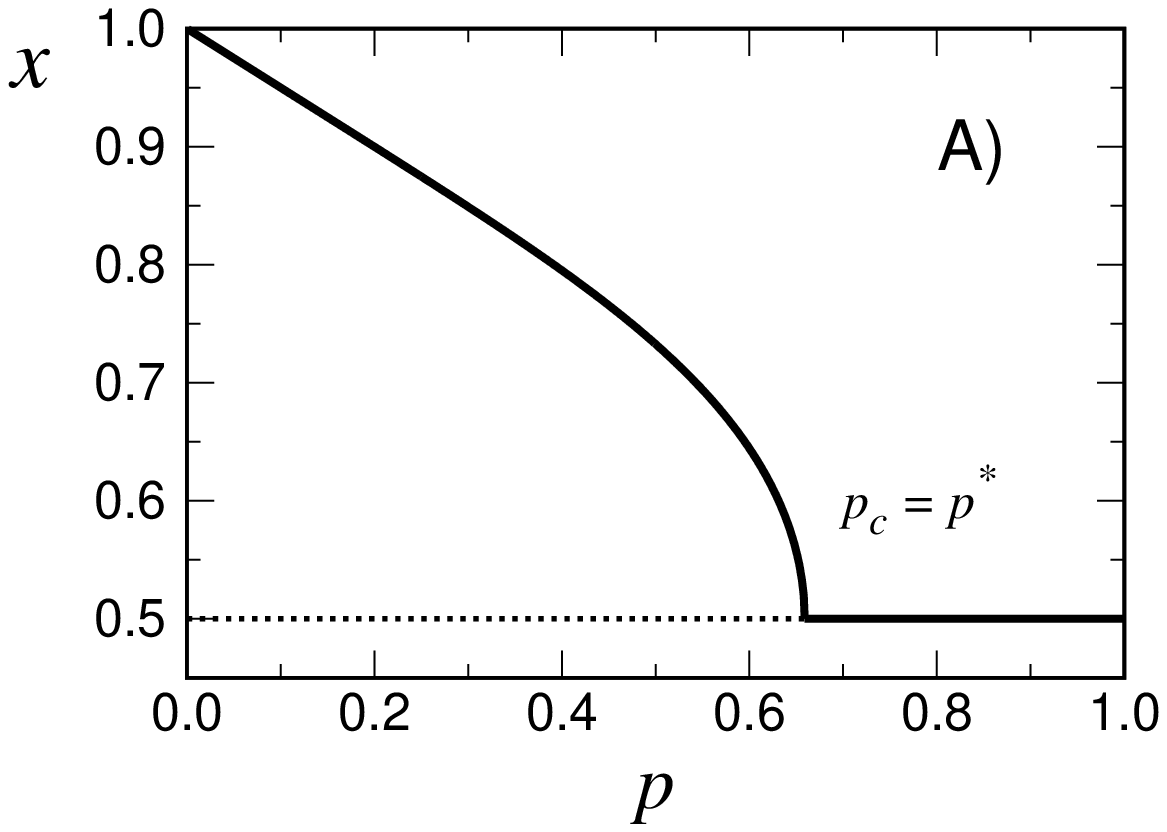} \\
\includegraphics[scale=0.46,angle=0]{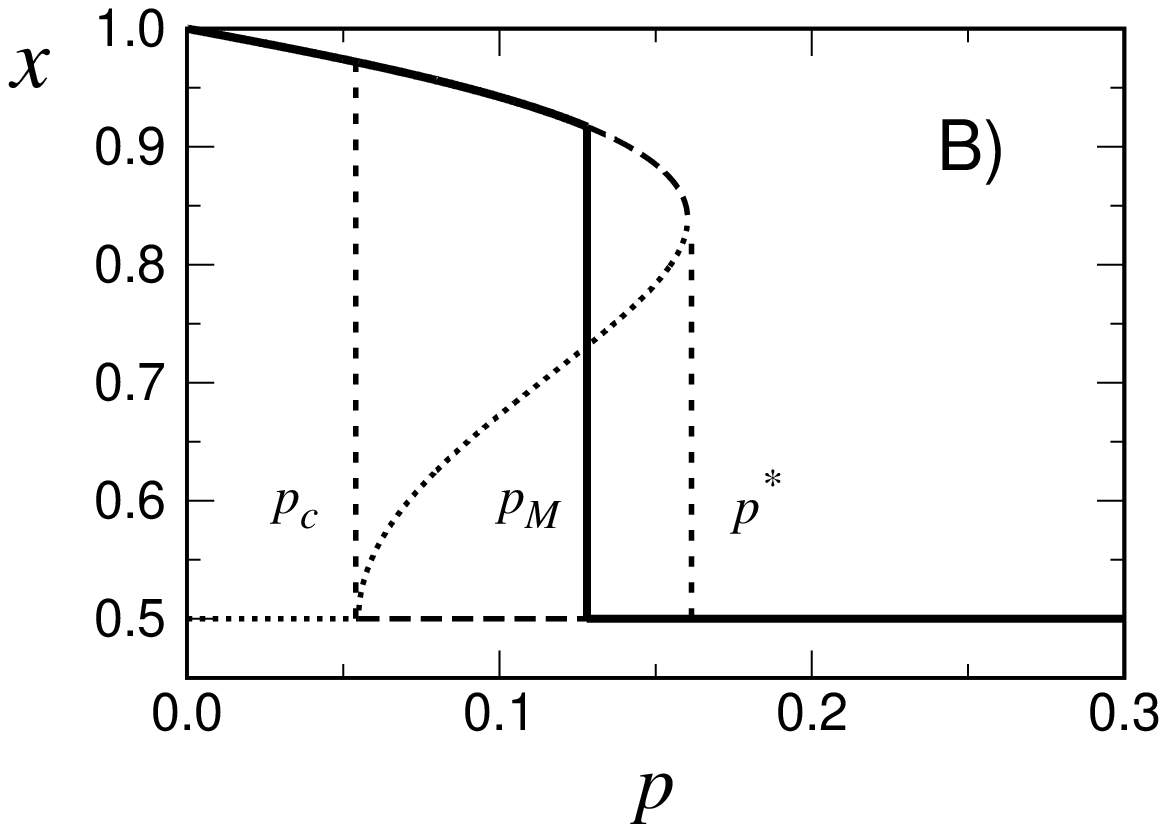} \\
\includegraphics[scale=0.46,angle=0]{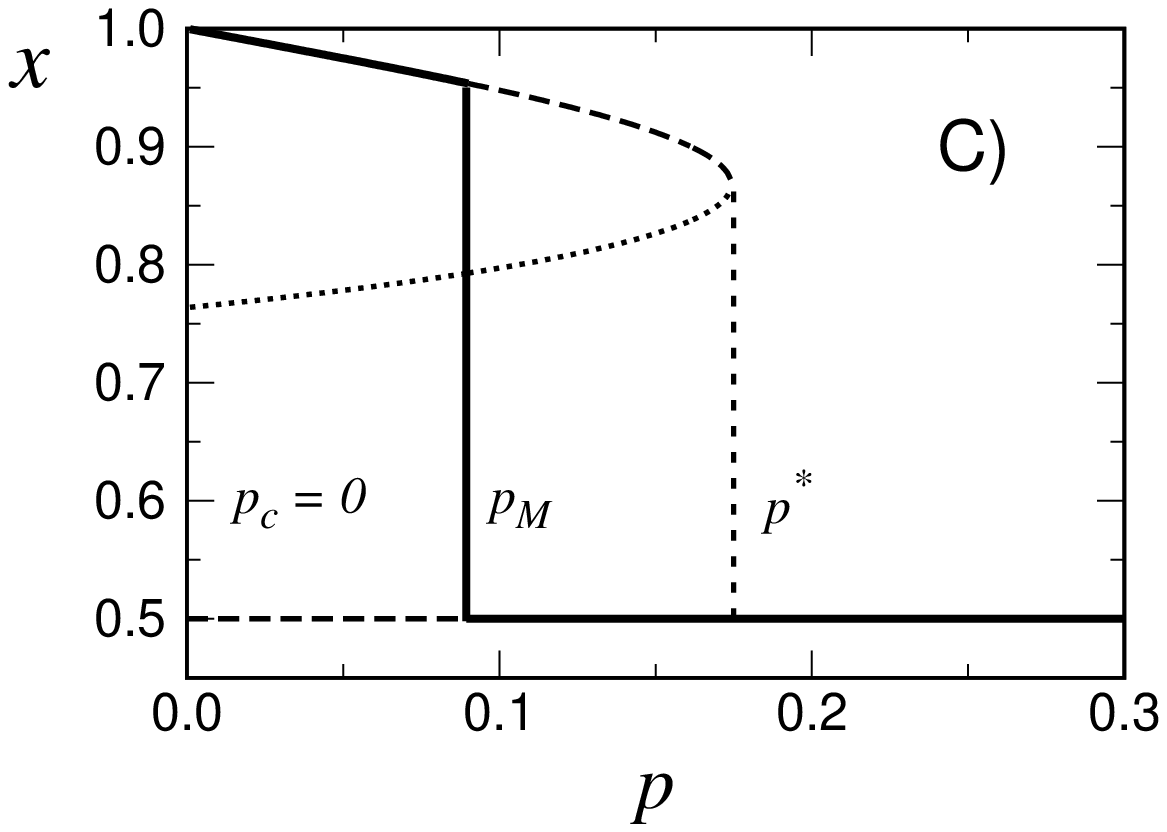} 
\end{center}
\caption{Schematic representation of the possible bifurcation diagrams of the model at the mean-field level. The black solid lines are the stable solutions, i.e. absolute minima of the potential $V(x)$ (macroscopic phases), dashed lines are metastable solutions (relative minima of the potential) and dotted lines are unstable solutions (maxima of the potential). The critical point (or lower limit of the hysteresis cycle), $p_c$, is given by Eq.~(\ref{pc_MF}) while the upper limit $p^*$ and the transition (Maxwell) point $p_\mathrm{M}$ are found numerically. Panel A) corresponds to a continuous phase transition occurring for $q_0\in(q_0^-,q_0^+)$. Panel B) corresponds to a first-order phase transition in the cases $q_0<q_0^-$, and $q_0>q_0^+$, and panel C) is a particular case of first-order transition occurring for $q_0<q_0^*$. The figure only shows the interval $x\in(1/2,1)$ and all lines have a mirror image in the interval $x\in(0,1/2)$, not shown for the sake of clarity. }
\label{fig:xst}
\end{figure}

Below $p<p_c$, $x=\frac12$ becomes unstable and two new stable solutions $x_\pm$, symmetric around $x=\frac12$, appear. These solutions correspond to ordered (O) phases such that the transition at $p=p_c$ between the order and disordered phases is continuous, see panel A) of Fig. \ref{fig:xst}. For some values of $(q,q_0)$, Eq.~(\ref{eq:extreme}) has, besides $x=\frac12$ and $x_\pm$, two additional real roots $x^*_\pm$ symmetric around $x=\frac12$ such that $x_-<x^*_-<1/2<x^*_+<x_+$. Again, the solutions $x_\pm$ are stable, while the $x^*_\pm$ are unstable. These four additional solutions appear in an interval $p_c<p<p^*$, being $p^*(q,q_0)$ a function of $q$ and $q_0$, while only the $x_\pm$ solutions exist for $p<p_c$. The value $p^*$ is found as the solution of $\upsilon'(x_\pm(p^*);p^* )=0$, an equation that has to be solved numerically. When several fixed points coexist, the potential $V(x)$ displays the absolute minimum at $x=1/2$ for $p\in(p_\mathrm{M},p^*)$ and at $x_\pm$ for $p\in(p_c,p_\mathrm{M})$. The first-order phase transition between the ordered and the disordered phases occurs at the Maxwell point $p_\mathrm{M}$ where the potential takes equal values $V(1/2)=V(x_\pm)$ at the minima, while $p_c$ and $p^*$ set the boundaries of the hysteresis cycle, see panel B) of Fig.~\ref{fig:xst}. Again, it does not seem to be possible to obtain an analytical expression for $p_\mathrm{M}$, and this value has to be obtained numerically.

The limiting condition $p_c=p^*$ necessary for the appearance of the ordered phases can be shown to be equivalent to $\upsilon'(1/2;q,q_0,p)=\upsilon'''(1/2;q,q_0,p)=0$. Eliminating $p$ from these equations one finds that the first-order phase transition only exists for $q_{0} < q_{0}^{-}$ and $q_{0} > q_{0}^{+}$, where the tricritical points are obtained as
\begin{equation} \label{q0c}
q_{0}^{\pm}(q) = \frac{1}{4} \left(5+2q\pm \sqrt{5+4q}\right).
\end{equation}

Take as an example, $q=12$ that yields $q_0^*\approx 4.75$, $q_{0}^{-} \approx 5.43$ and $q_{0}^{+} \approx 9.07$ indicating that first-order transitions occur for $q_0<q_0^-$ or $q_0>q_0^+$, namely $q_0=2,3,4,5,10,11,12$, while continuous transitions occur for the other values $q_0=6,7,8,9$. For $q_0<q_0^*$, namely $q_0=2,3,4$, the disordered phase is either stable or metastable, but never unstable. For $q=6$, continuous transitions occur for $q_0=3,4,5$ while first-order transitions occur for $q_0=2,6$. In the case $q_{0}=1$, the only observed fixed point is $x=1/2$ (stable) which presents no transitions as a function of $p$.

\begin{figure}[h!]
\begin{center}
(a) Mean-field (MF)\\
\includegraphics[scale=0.15,angle=-90]{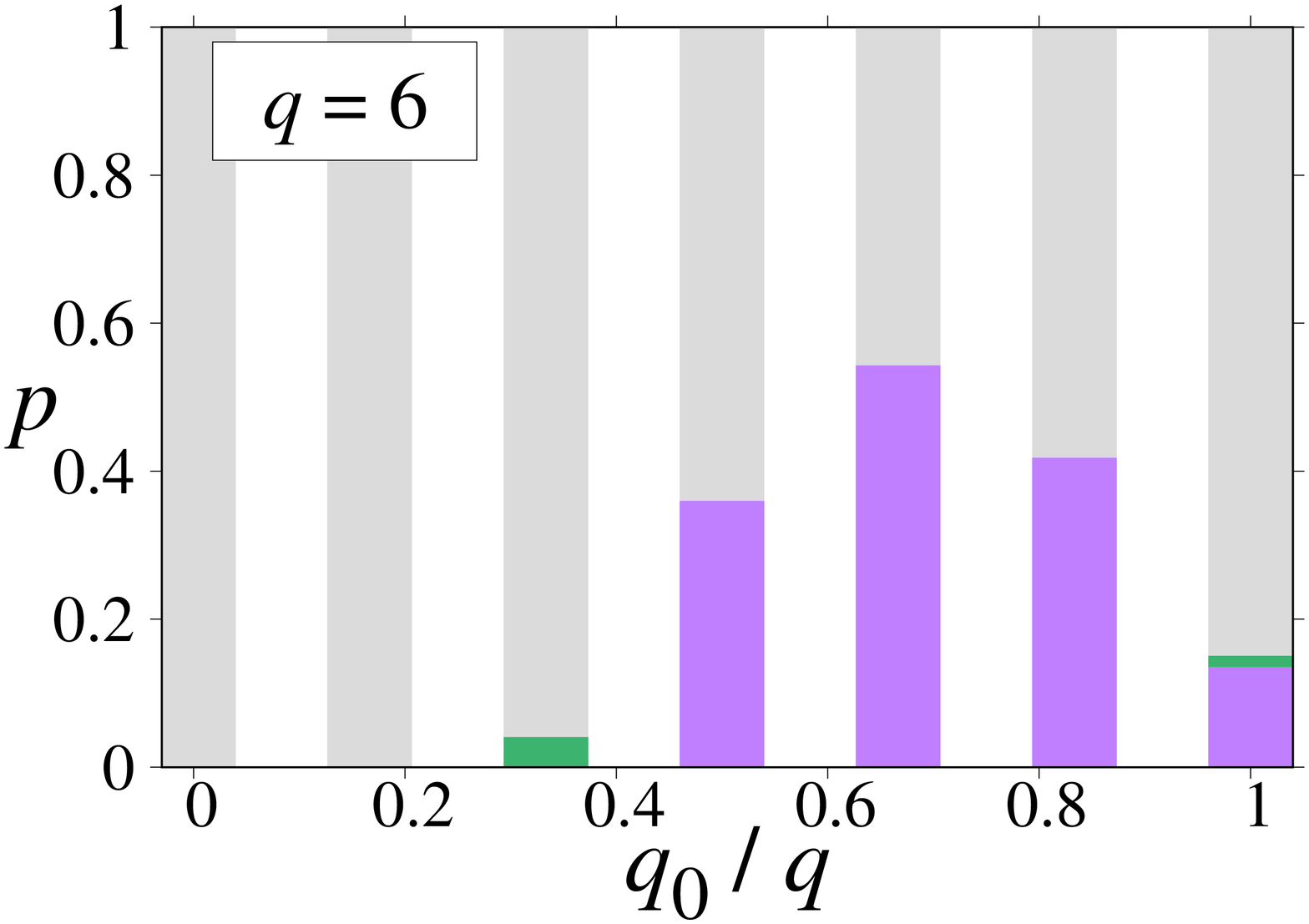}
\includegraphics[scale=0.15,angle=-90]{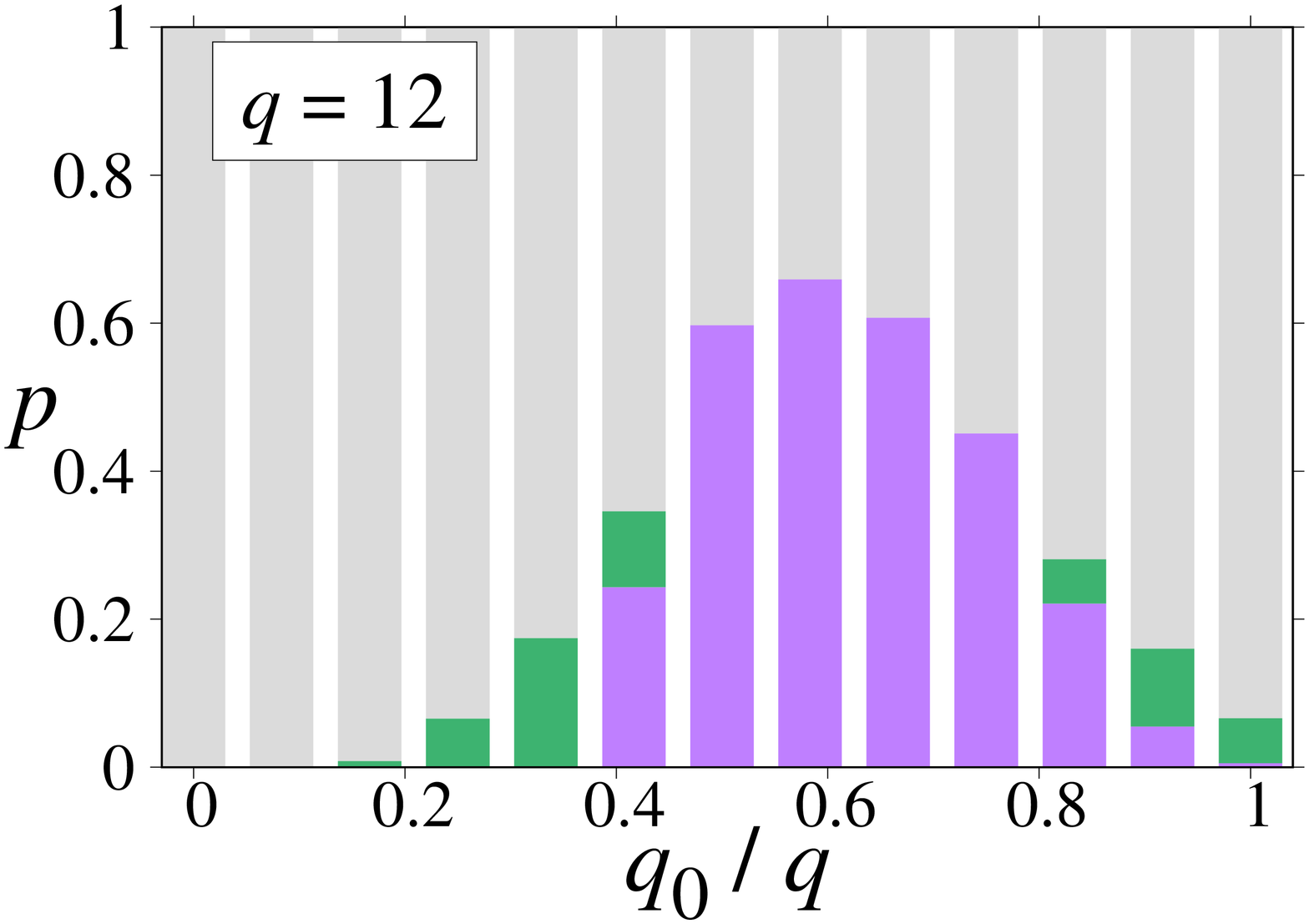} 
\end{center}
(b) Without repetition, in random regular networks\\
\includegraphics[scale=0.15,angle=-90]{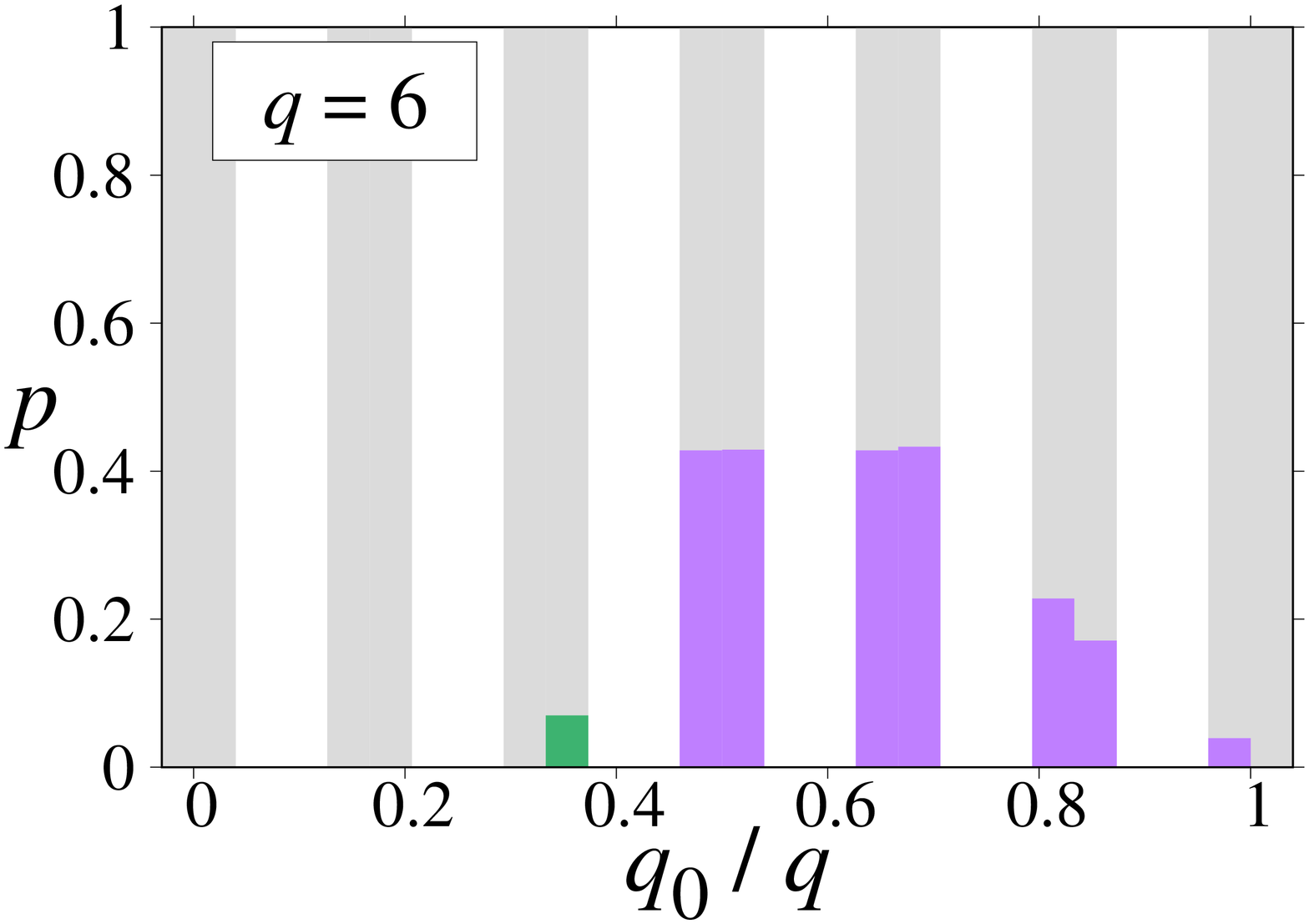}
\includegraphics[scale=0.15,angle=-90]{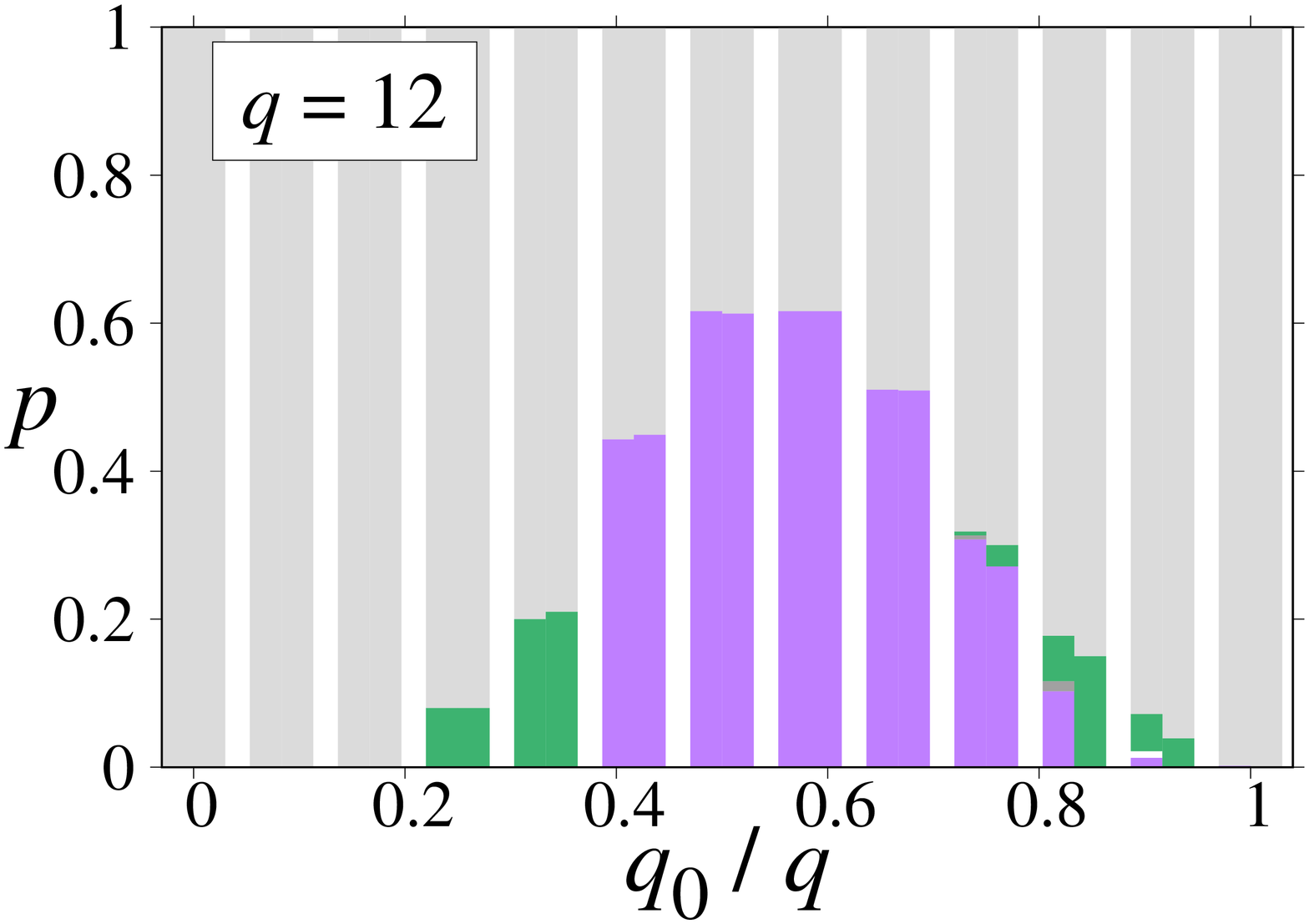}\\[2mm]
(c) With repetition, in random regular networks\\
\includegraphics[scale=0.15,angle=-90]{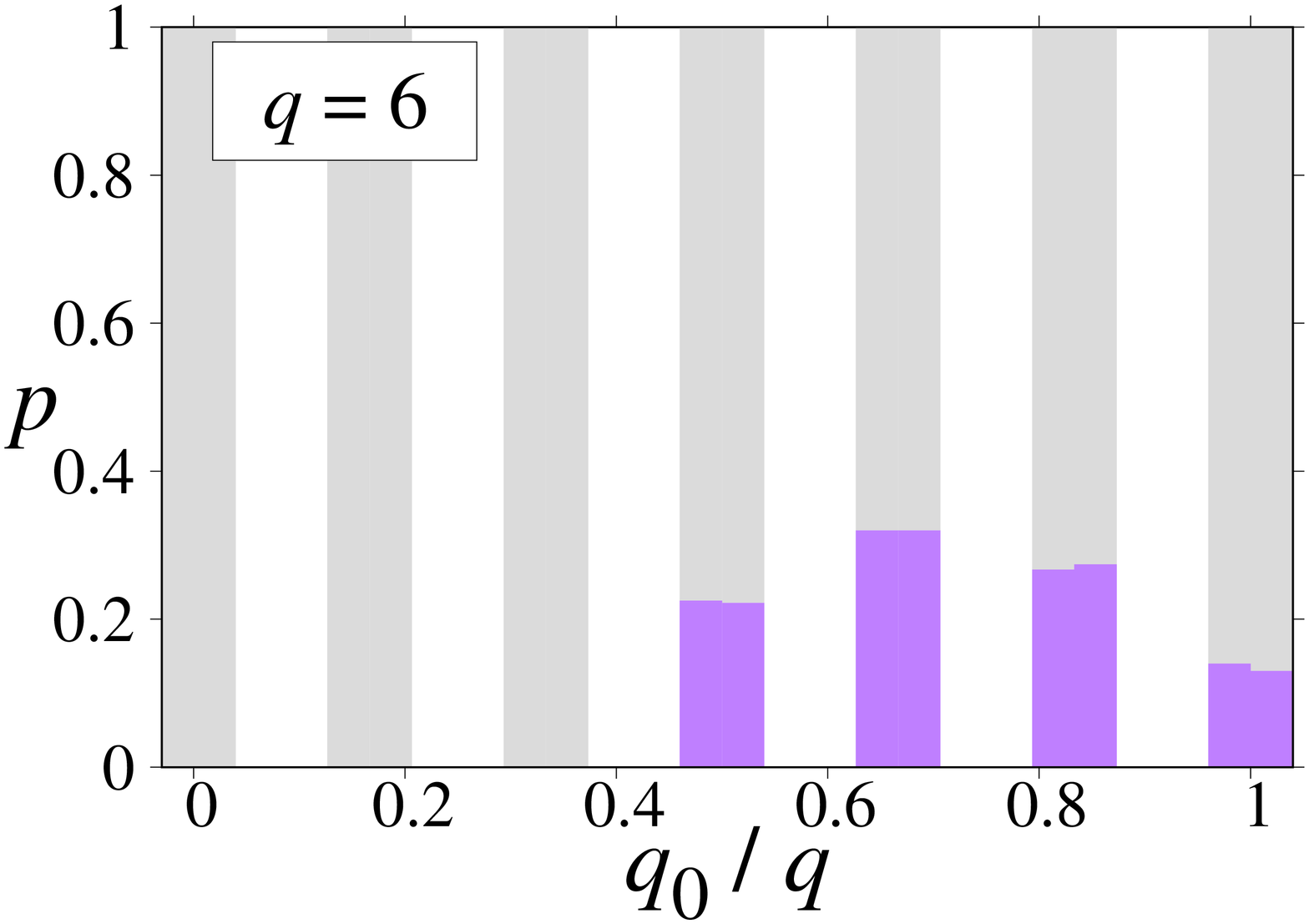}
\includegraphics[scale=0.15,angle=-90]{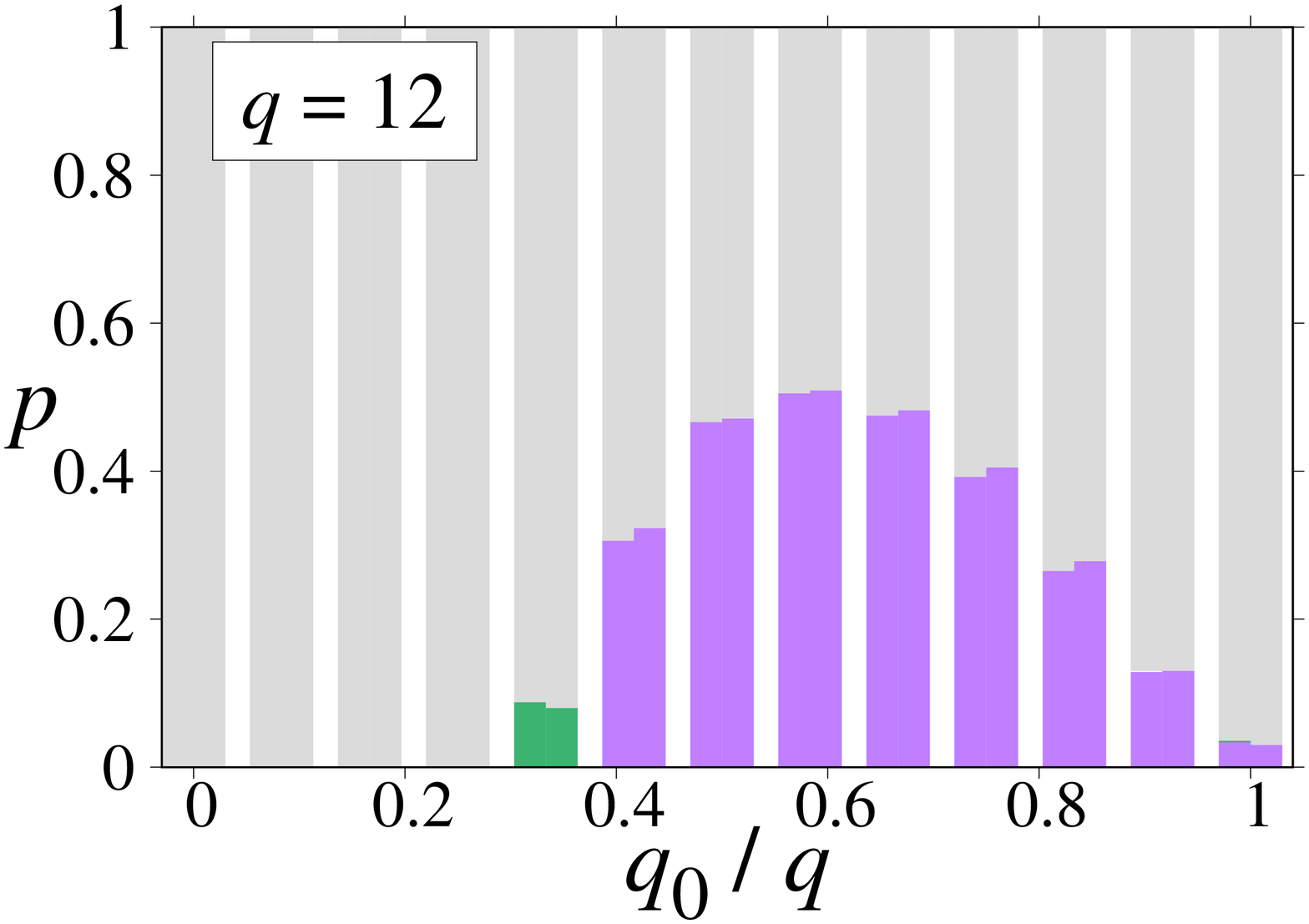} \\[2mm]
\caption{
Panel (a) shows mean-field phase diagrams $p$ versus $q_{0}/q$ for two different values of $q$. Panels (b,c) show the equivalent results for the random regular network again for two different values of $\mu=q$ ($\mu$ being the mean degree of the network, see section \ref{sec:PA}). In the figures, each phase is represented by a different color and pattern:
disordered (D, light-gray), ordered (O, lilac) and ordered-disordered (OD, green).
Transition O-D is continuous while the other ones are discontinuous (indicated by 
black segments).
In panels (b,c) for each value of $q_0/q$, the left (resp., right) half of the bars corresponds to the pair approximation (resp., numerical simulations). 
} 
\label{fig:diagrams}
\end{figure}

An alternative representation of these results is presented in the phase diagrams depicted in panel (a) of Fig.~\ref{fig:diagrams} for $q=6$ (left) and $q=12$ (right). Gray regions are disordered phases where $x=1/2$ is the only stable solution. The lilac region corresponds to the existence of stable symmetric fixed points $x_\pm$, while in the green region $x=1/2$ and $x_\pm$ coexist as stable fixed points. The upper limit of the lilac region is the value $p_c$ and the upper limit of the green region is the value $p^*$. If the lilac region goes directly into the gray, it indicates a continuous transition between order and disordered phases. When the green region goes into gray, it indicates a first-order phase transition occurring at the Maxwell point (not depicted) somewhere in the green region. 

\section{Pair approximation }
\label{sec:PA}

Now, we wish to extend the results of the previous section to more complex network topologies. To this end, let us consider an arbitrary network with $N$ nodes where each node has a different number of connections, or degree, $k$ and we denote by $P(k)$ the degree distribution. An active link is defined as joining two nodes in different states. While in the all-to-all connected network the fraction of active links $\rho$ is related to the density $x$ of nodes in the $\oplus$ state by $\rho=2x(1-x)$, in a general network $\rho(t)$ and $x(t)$ must be considered as independent variables for which one needs to write down evolution equations. The pair approximation (henceforth, PA) is one of the simplest, yet very successful approaches, to derive such a set of equations taking into consideration the underlying network structure. 
One of the main differences between the all-to-all connectivity and a non-fully connected network is that in the latter the flipping probability depends on whether in the random selection of the $q$ neighbors one allows or not for repetitions of the chosen nodes. Recall that in the large $N$ limit, this difference was irrelevant in the all-to-all connectivity as the probability of repetition was negligible. If the chosen node has $k$ links amongst which $\ell\in(0,k)$ of them are active, the probability that this node flips its state is:
\begin{equation}
 \label{eq:F}
F(\ell;k,q,q_0,p)=(1-p)f(\ell;k,q,q_0)+p/2,
\end{equation}
where 
\begin{equation} 
f(\ell;k,q,q_0) \equiv 
\left \{ \begin{array}{lr}
 \displaystyle \sum_{j = q_0}^{q} {{q}\choose{j}}{{k-q}\choose{\ell-j}}\left/{{k}\choose{\ell}}\right., &\text{(a)} \\[5mm]
\displaystyle \sum_{j = q_0}^{q} {{q}\choose{j}} \, 
 \left(\frac{\ell}{k} \right)^j \left(1-\frac{\ell}{k}\right)^{q-j}, &\text{(b)} 
\end{array}
\right.
\label{eq:f}
\end{equation} 
depending on whether repetition in the selection is forbidden, case (a), or allowed, case (b). In the former case, it is understood that $f(\ell;k,q,q_0)=0$ if $k<q$. The combinatorial number in (a) is also understood to be equal to zero if $\ell<j$.

We focus in this work in a version of the PA based on the single degree distribution $P(k)$ as developed in Refs.~\cite{vazquezNJP,vazquezJST} and in more detail in Ref.~\cite{antonioNJP}. This is expected to work well for random networks that are not highly clustered nor correlated, otherwise a heterogeneous version of the PA considering joint degree distributions should be used. In this approach, we consider as description variables the fraction of nodes $x_{k}$ with degree $k$ that hold $\oplus$ opinions, and the fraction $\rho$ of active links. Following Ref. \cite{antonioNJP}, the rate equations for the description variables $\lbrace x_{k}, \rho \rbrace$ can be derived as follows:
\begin{eqnarray}
\notag
\frac{d\rho}{dt} &=& \frac{2}{\mu}\sum_k\sum_{i=\oplus,\ominus} P(k) P_{i,k} \langle (k-2\ell)F(\ell;k,q,q_0,p) \rangle_{\rho_i},\\
\label{eq:rho_peralta}\\
\label{eq:x_peralta}
\frac{dx_{k}}{dt} &=& -\sum_{i=\oplus,\ominus} S_i P(k) P_{i,k} \langle F(\ell;k,q,q_0,p) \rangle_{\rho_i},
\end{eqnarray}
where $S_\oplus=1$, $S_\ominus=-1$, $\mu \equiv \sum_{k} P(k) k$ is the mean degree and $\langle \cdots \rangle_{\rho_i}$ is the average calculated over the binomial probability ${k\choose \ell}\rho_i^\ell(1-\rho_i)^{k-\ell}$. Moreover, we identify $P_{\oplus,k}=x_{k}$, $P_{\ominus,k}=1-x_{k}$, and $\rho_\oplus=P(\ominus|\oplus)=\rho/(2 x_{L})$, $\rho_\ominus=P(\oplus|\ominus)=\rho/[2(1-x_{L})]$, where $x_{L}=\sum_{k} P(k) k x_{k}/\mu$ is the so-called {\slshape link-magnetization}, and $P(i'|i)$ is the conditional probability of selecting a neighbor with opinion $i'$ if the node chosen for updating holds opinion $i$. 

It is straightforward to show that a particular solution of Eqs. (\ref{eq:x_peralta}) is $x_{k}(t) = x_{L}(t)$ for all degrees $k$. As shown in \cite{antonioNJP} the use of this particular solution is appropriate as far as we restrict the analysis to steady-state deterministic values. Using this simplification it is possible to reduce the problem to two closed rate equations for the density of active links $\rho(t)$ and the density of nodes in the $\oplus$ state, $x(t)$, namely:
\begin{eqnarray}
\frac{d\rho}{dt} &=& \frac{2}{\mu}\sum_k\sum_{i=\oplus,\ominus} 
P(k)\, P_{i} \langle (k-2\ell)F(\ell;k,q,q_0,p) \rangle_{\rho_i},\notag\\
\label{eq:rho}\\
\label{eq:x}
\frac{dx}{dt} &=& -\sum_k\sum_{i=\oplus,\ominus} \frac{k}{\mu} P(k)\,S_i\,P_{i} \langle F(\ell;k,q,q_0,p) \rangle_{\rho_i},
\end{eqnarray}
where $P_{\oplus}=x$, $P_{\ominus}=1-x$, $\rho_\oplus=\rho/(2 x)$, $\rho_\ominus=\rho/[2(1-x)]$. We now determine the fixed points of these equations and their stability for different values of $q$ and $q_0$ in the cases with and without repetition in the selection of the neighbors.

\subsection{Without repetition} 

In this case, we replace Eqs.~(\ref{eq:F},\ref{eq:f}a) into the rate equations (\ref{eq:rho},\ref{eq:x}). It turns out that, due to some cancellation of the combinatorial numbers, the averages $\langle \dots\rangle_{\rho_i}$ over the binomial distributions lead to relatively simple expressions which are linear in $k$, see Appendix A for details.
Therefore, the further average over the degree distribution $\sum_k P(k)$ leads to expressions that depend only on the first moment $\langle k \rangle \equiv \mu$, namely
\begin{eqnarray} 
&&\frac{d \rho}{dt} = p(1-2\rho) +\label{eq:rho_without}\\
&&+ \frac{2(1-p)}{\mu} \biggl[ 
(1-x) G_2(\rho_\ominus;q,q_0,\mu)+xG_2(\rho_\oplus;q,q_0,\mu)\biggl], \nonumber \\ 
&& \frac{dx}{dt} = (1-x)G(\rho_\ominus;q,q_0,p) - 
x G(\rho_\oplus;q,q_0,p) \, .
\label{eq:without}
\end{eqnarray}
Where the function $G$ has been defined in Eqs.~(\ref{eq:G})-(\ref{eq:gx}) and $G_2$ is defined in Appendix A.

 It can be easily checked that $x(p)=1/2$ (the disordered phase) is always a fixed point of Eqs. (\ref{eq:rho_without},\ref{eq:without}) for any value of $p,\,q,\,q_0$. This is a straightforward consequence of the symmetry of the rates around $x=1/2$. The corresponding fixed point $\rho(p)$, obtained from $\sum_kP(k)\langle (k-2\ell)F(\ell;k,q,q_0,p)\rangle_{\rho(p)}=0$, can not be found analytically for general $(p,q,q_0)$. However, at the special point $p_c(q,q_0)$ where the fixed point $x=1/2$ loses its stability, a general argument \cite{gleeson} valid for any model with up-down symmetric rates in random regular networks defined by $P(k)=\delta(k-\mu)$, leads to $\rho_c\equiv\rho(p_c)=\frac{\mu-2}{2(\mu-1)}$. The condition to obtain $p_c$ is thus reduced to the single equation 
$\langle (\mu-2\ell)F(\ell; \mu, q, q_0, p_c)\rangle_{\rho_c} = 0$. 
This leads to the analytical result:
\begin{equation} \label{eq:pc}
p_c^{-1} = 
1+\dfrac{2^{q-1} \bigl(\frac{\mu}{\mu-2}\bigr)^{q_0}\bigl(\frac{\mu-1}{\mu}\bigr)^{q} 
\Gamma(q_0+1)\Gamma(q - q_0 +1) }
{\Gamma(q+1) [q_0- {_2F_1}(1, q_0-q, q_0+1,2/\mu -1)]},
\end{equation}
that generalizes the result for the $q$-voter model~\cite{qvoterPA} 
$p_c^{-1} = 1+ \frac{2^{q-1}}{q-1} (\frac{\mu-1}{\mu-2})^q$ and for 
the MF Eq.~(\ref{pc_MF}), recovered in the limit $\mu\to\infty$.
 Since for the case without repetition the rate equations (\ref{eq:without}) depend only on the average degree, the values of $p_c$ and $\rho_c$ are also valid for networks with an arbitrary degree distribution.

\begin{figure}[h!]
\begin{center}
\includegraphics[scale=0.47,angle=0]{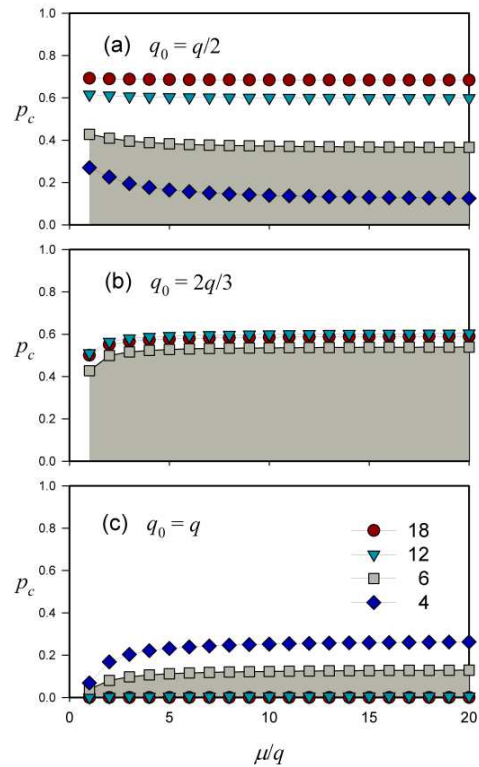} 
\end{center}
\caption{Critical point $p_c$ vs. $\mu/q$ predicted by Eq.~(\ref{eq:pc}), in networks with average connectivity $\mu$, for (a) $q_0=q/2$, (b) $q_0=2q/3$ and (c) $q_0=q$, without repetition. Each symbol/color corresponds to a different value of $q$. The shaded region emphasizes the region where the ordered state (in either O or OD phases) is stable when $q=6$ as a reference. 
}
\label{fig:keffect} 
\end{figure}

Figure~\ref{fig:keffect} shows the effect of increasing $\mu$, 
predicted by Eq.~(\ref{eq:pc}). 
The examples correspond to 
(a) simple majority ($q_0=q/2$), 
(b) two-thirds majority or {\slshape byzantine agreement} ($q_0=2q/3$), 
and 
(c) the $q$-voter ($q_0=q$). 
In case (a), the quantity of elements in the influence group with opposite opinions 
is equal or greater than the quantity of elements sharing the central node opinion, 
in case (b) the opposite opinions are at least twice, 
while in (c) total consensus of opposite opinions is required. 
  
We first note that to attain an ordered collective state where one of the opinions dominates 
is facilitated in cases (a) and (b) with respect to the $q$-voter model (c). 
Moreover, the two-thirds condition is optimal for small size $q$. 
For increasing $q$, dominance of one of the opinions 
is facilitated in case (a) and degraded in case (c) while (b) is more 
robust against changes in group size. 
In (b) and (c), increasing the relative connectivity $\mu/q$ favors order, 
but the opposite effect is observed in (a). 
For $\mu/q\to \infty$, $p_c$ tends to the MF value in all cases, as expected. 

\begin{figure}[h!]
\begin{center}
\includegraphics[scale=0.43,angle=0]{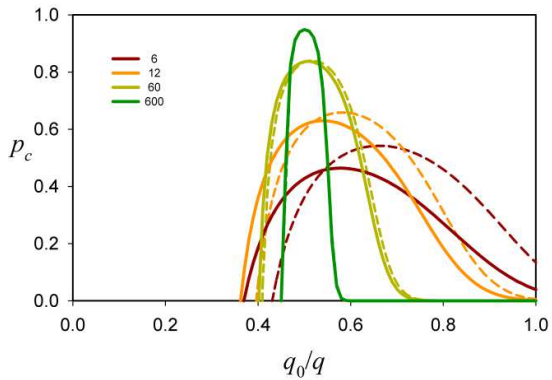} 
\end{center}
\caption{Critical point $p_c$ vs. $q_0/q$, predicted by Eq.~(\ref{eq:pc}), 
in networks with average connectivity $\mu/q=1$ (solid lines) 
and mean field limit (dashed lines). The system orders below the curves. 
Values of $q$ are indicated in the legend.
}
\label{fig:q0effect} 
\end{figure}

The level of the threshold $q_0$ appears to have a larger impact 
on the collective state of the system than the structure parameter $\mu$. 
The effect of $q_0$ predicted by Eq.~(\ref{eq:pc}) is shown in Fig.~\ref{fig:q0effect}, 
where the critical curves (below which the system orders) 
are given for the extreme cases $\mu=q$ (solid lines) 
and MF limit (dashed lines) given by Eq.~(\ref{pc_MF}). 
For increasing $q$, both critical curves tend to collapse, and 
the optimal value of $q_0$ to attain ordered configurations is shifted from $q_0\simeq 2q/3$ towards $q_0=q/2$, as observed also in Fig.~\ref{fig:keffect}. 
Moreover, order occurs for increasing level of independence $p$, 
but becomes restricted to the vicinity of $q_0/q=1/2$.

Additional stable fixed points of the rate equations (\ref{eq:rho_without},\ref{eq:without}) where $x\ne 1/2$ (ordered phases) can appear depending on the value of the parameters $(p,q,q_0)$ and lead to first-order phase transitions for $q_0<q_0^-$ and $q_0>q_0^+$, in a similar fashion to the one depicted in Fig. \ref{fig:xst}. It is possible to obtain an equivalent expression of $q_{0}^{\pm}$ as Eq.~(\ref{q0c}) in the case of networks, but an additional approximation is needed in Eq.~(\ref{eq:without}). Following the reasoning of Refs. \cite{vazquezNJP,vazquezJST,antonioChaos,antonioNJP}, at the critical point $p=p_{c}$ we can eliminate the $\rho$ variable as follows $\rho \approx 4 \rho_{c} x (1-x)$. This leads to a closed equation $\frac{dx}{dt}=\tilde{\upsilon}(x;q,q_0,p)$ for the variable $x$ that can be analyzed in the same way than in the mean-field scenario. The result is that the limits for the existence of first-order transitions are:
\begin{align} \label{q0c_net}
& q_{0}^{\pm}(q,\mu) \approx \frac{1}{2} \bigg[6+(2q-7)\rho_{c} \bigg.\notag\\
&\bigg. \pm \sqrt{16+\rho_{c} \left(-44+4 q (1-\rho_{c})+29 \rho_{c} \right)} \bigg].
\end{align}

\begin{figure}[h!]
\begin{center} 
\includegraphics[scale=0.26,angle=-90]{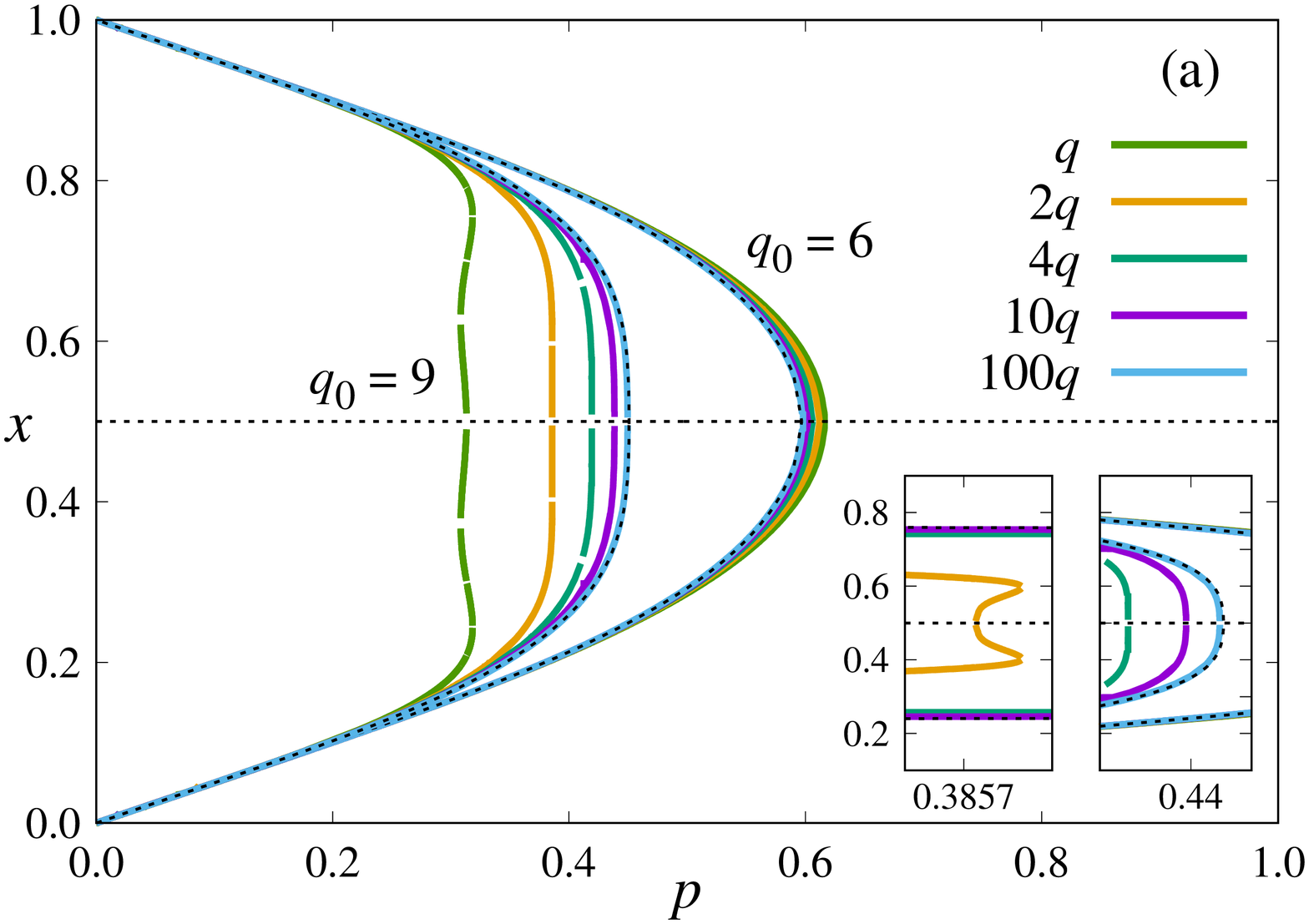} \\
\includegraphics[scale=0.26,angle=-90]{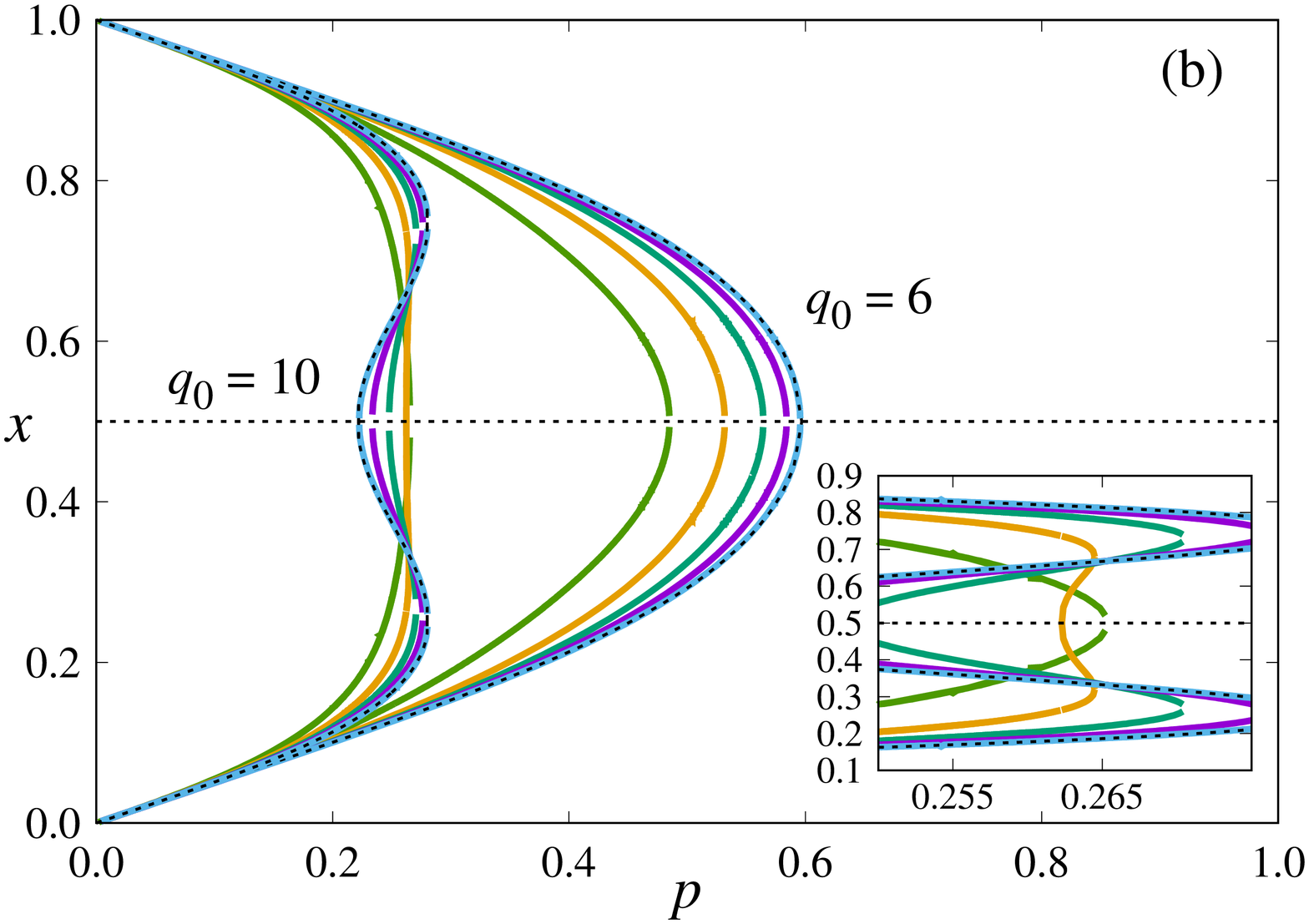} 
\end{center}
\caption{Stability diagrams $x$ vs. $p$, for $q=12$ and two values of $q_0$. 
Black dashed lines represent the mean-field solution. 
Colored curves are predicted by the PA for average connectivity $\mu$ (indicated in the legend 
as multiples of $q$): 
(a) without repetition for any random network with average connectivity $\mu$, 
(b) with repetition for random regular networks, i.e. $P(k)=\delta(k-\mu)$. 
The insets show zooms of the main frame. 
}
\label{fig:MFlimit}
\end{figure}

In the case $\mu=q=12$, Eq.~(\ref{q0c_net}) yields $q_{0}^{-} \approx 5.00$ and $q_{0}^{+} \approx 8.73$. Therefore a discontinuous transition is predicted for $q_0=9$, contrarily to the mean-field case (equivalent to $\mu\to\infty$) where the transition at $q_0=9$ was continuous. The change of nature of the transition from continuous to discontinuous when decreasing the value of $\mu$ occurring for $q_0=9$ is shown in the upper panel of Fig.~\ref{fig:MFlimit}. Note that, additionally to the MF-like phenomenology (where one, three or five fixed point solutions for $x$ were obtained), it is found that in the case with repetition up to seven fixed points (out of which four are stable) can be found in a narrow range of parameters, see, e.g. Fig. \ref{fig:MFlimit} for $q_0=9$ and $\mu=q$. However, these multistability solutions with four stable points appear to be spurious as they are not found in the numerical simulations as we will show in the next sections. Other features of the phase diagram will be discussed in Section \ref{sec:results} in the context of the comparison with numerical simulations.

Finally, let us mention that for $\mu\gg q$, we can use the mean-field relation $\rho=2x(1-x)$ to show that both the rate equation (\ref{eq:without}) for $x$ and the expression (\ref{q0c_net}) for the limits for the existence of discontinuous transitions, reduce to the corresponding expressions in the mean-field limit Eqs. (\ref{eq:drift}) and (\ref{q0c}), respectively.

\subsection{With repetition} 
 
When repetition in the choosing of the $q$ neighbors is allowed, we must use, inside the rate Eqs.~(\ref{eq:x}), 
the corresponding definition of $f(\ell;k,q,q_0)$ given in Eq.~(\ref{eq:f}b). For this modality, we did not manage to find general closed expressions for the averages $\langle \dots\rangle_{\rho_i}$ over the binomial distribution. Moreover, while without repetition the binomial averages depend only on the average degree $\mu$, 
independently of other features of the degree distribution $P(k)$, 
in the variant with repetition, also negative moments 
$\langle k^{-m} \rangle$, with $1 \le m \le q-1$, contribute. It is still true that $x=1/2$ is a fixed point and that, as in the case without repetition, in the limit $\mu \gg q$, it is $\rho=2x(1-x)$ and the rate equation for $x$ becomes 
Eq.~(\ref{eq:drift}). However, in general, the whole analysis of the existence and stability of the fixed points must be performed numerically. A difficulty of the numerical calculation is that the binomial coefficients appearing in the rate equations take huge values for large $k$, while $\rho^{k}$ takes very small ones, leading to a very delicate numerical evaluation. To circumvent this 
problem when solving numerically equations ((\ref{eq:rho},\ref{eq:x})) in order to find the fixed 
points~\cite{software}, 
we made expansions using the moments of the binomial distribution and the negative moments of the degree distribution, instead of evaluating the bare expressions.

Before making a detailed comparison of the theoretical predictions of the pair approximation with numerical simulations for different underlying networks, let us stress the main differences that appear when allowing or not repetition in the selection of the $q$ neighbors in the influence group.

\subsection{Comparison ``with vs. without'' repetition} 

We have found numerically the fixed points of the rate equations (\ref{eq:rho},\ref{eq:x}) in order to obtain the stability diagrams for networks of increasing connectivity $\mu=m\,q$ (with $m \in \mathbb{N}$ and $q=12$), 
predicted by the PA, for the rules (a) without and (b) with repetition. The results are displayed in Fig.~\ref{fig:MFlimit}, 
together with the mean-field results discussed above. 
In the case without repetition, PA calculations depend only on the average 
degree $\mu$. With repetition, the PA depends on other moments, then 
we choose for this example $P(k)=\delta(k-\mu)$, which is the distribution of random regular networks to be considered in detail in another section. 
The values of $\mu$ are indicated in the figure as multiples of $q$. 

Values of $q_0$ were selected to illustrate qualitatively different behaviors. 
For $q_0\simeq q/2$, the order/disorder transition is typically 
continuous, while for $q_0\simeq q$, it is typically discontinuous. 
In all cases, the MF result is analytically recovered in the limit of very large $\mu$ 
and achieved in the simulations performed (with or without repetition) 
in fully connected networks (not shown). 
In most cases, the nature of the transitions is not altered when changing $\mu$, 
except for some cases, as those displayed in the figure. 

(a) Without repetition, the PA prediction remains close to the MF limit when $q_0\simeq q/2$, even for small $\mu$. 
Note that the critical point slightly increases when $\mu$ decreases, 
meaning that, surprisingly, the network favors order with respect to the MF. 
The deviation from MF is more pronounced when $q_0$ approaches $q$. 
Even the nature of the transition changes from continuous to discontinuous when departing 
from the MF (bear in mind that the multistability of the case $q_0=9$ around $p\simeq 0.3$ is spurious, 
absent in simulations although the discontinuous character is still observed, 
as we will see in Sec. \ref{sec:results}.)

(b) Differently, with repetition, a noticeable deviation from MF occurs for $q_0\simeq q/2$. 
The critical point increases with $\mu$ but the transition remains always continuous. 
When $q_0\simeq q$, although the variation of the critical point is smaller, there is 
a change in the nature of the transition, from discontinuous to continuous, when $\mu$ 
decreases (at $q<\mu <2q$), contrarily to the case without repetition.

\section{Comparison with numerical simulations}
\label{sec:results}

Simulations of the noisy threshold $q$-voter opinion dynamical rules were extensively performed. In the following, we compare the predictions of the pair approximation with the numerical simulations over different topologies of the network of contacts. We focus on the case $\mu=q$, which is the one more distant from the MF limit, when repetitions are forbidden.

\subsection{Random regular networks}

\begin{figure*}[h!]
\begin{center}
\includegraphics[scale=0.23,angle=-90]{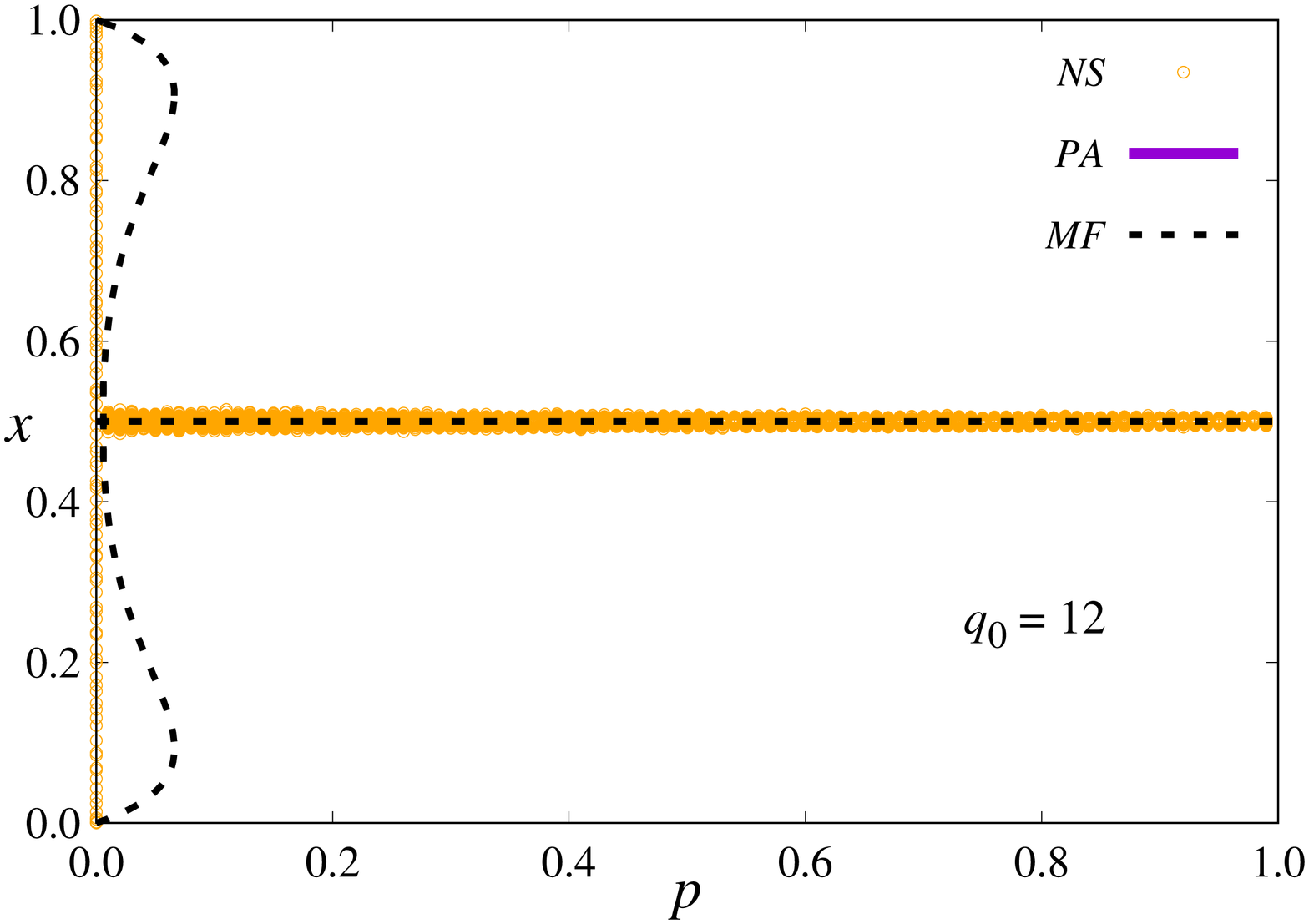} 
\includegraphics[scale=0.23,angle=-90]{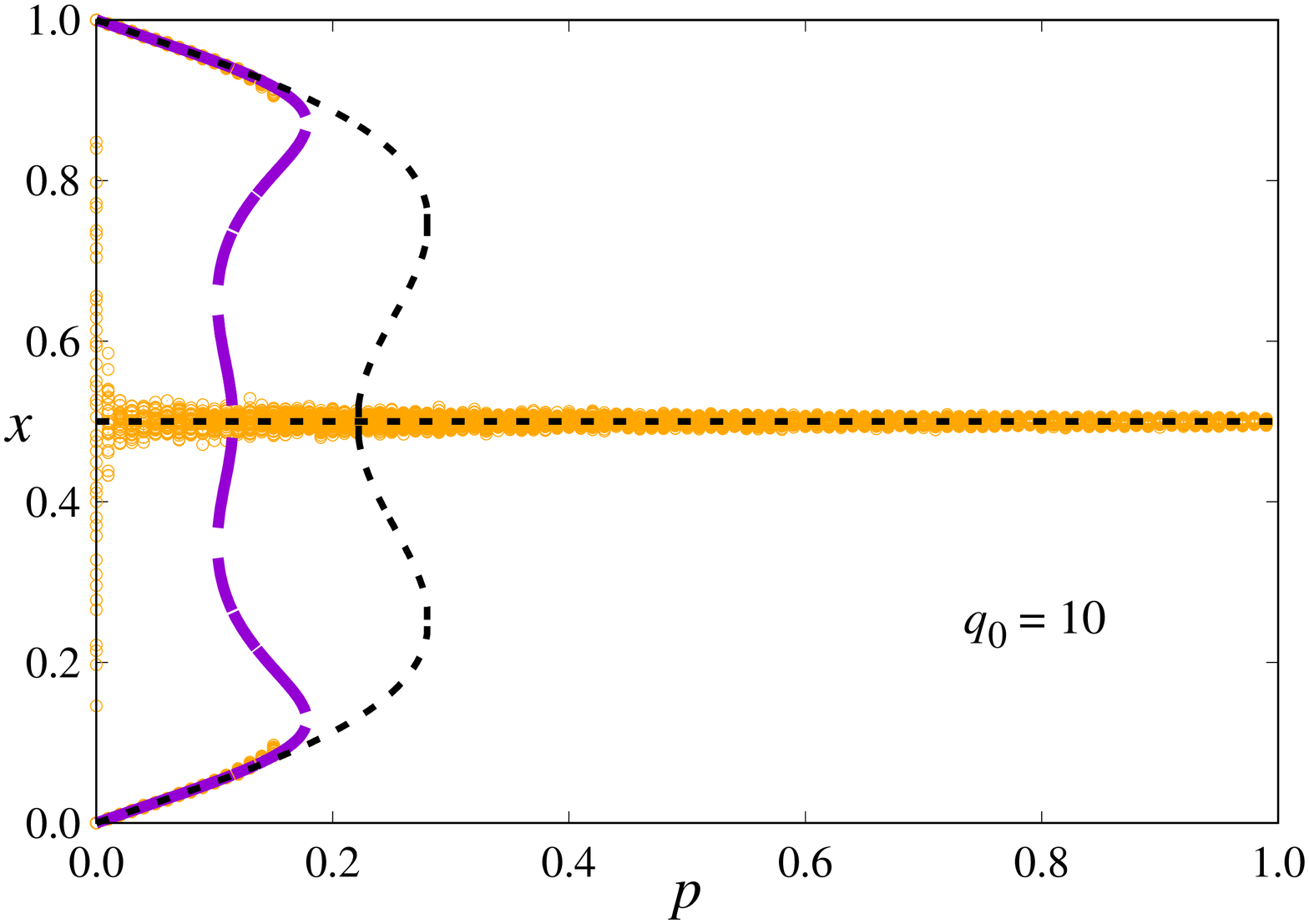} \\
\includegraphics[scale=0.23,angle=-90]{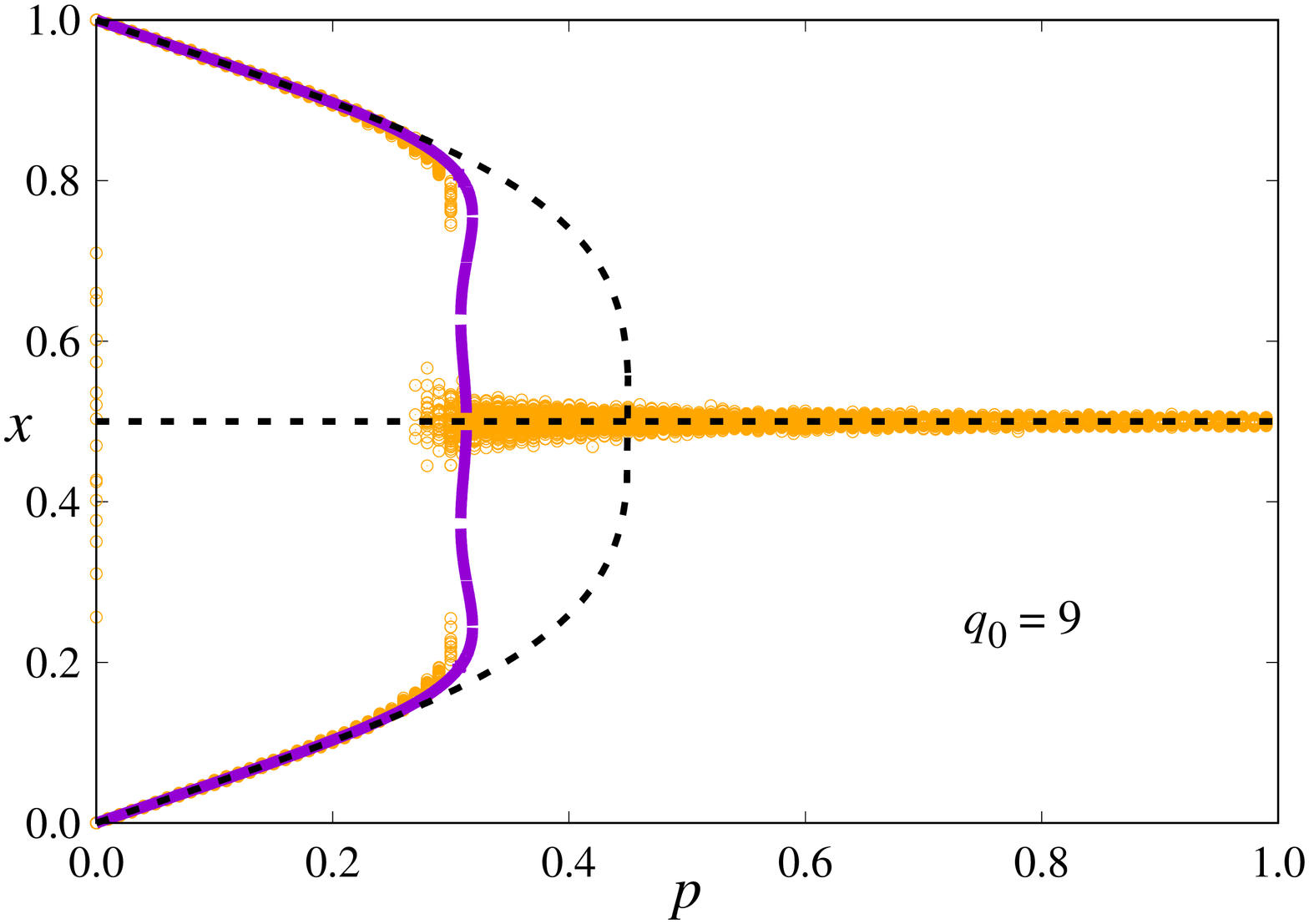}
\includegraphics[scale=0.23,angle=-90]{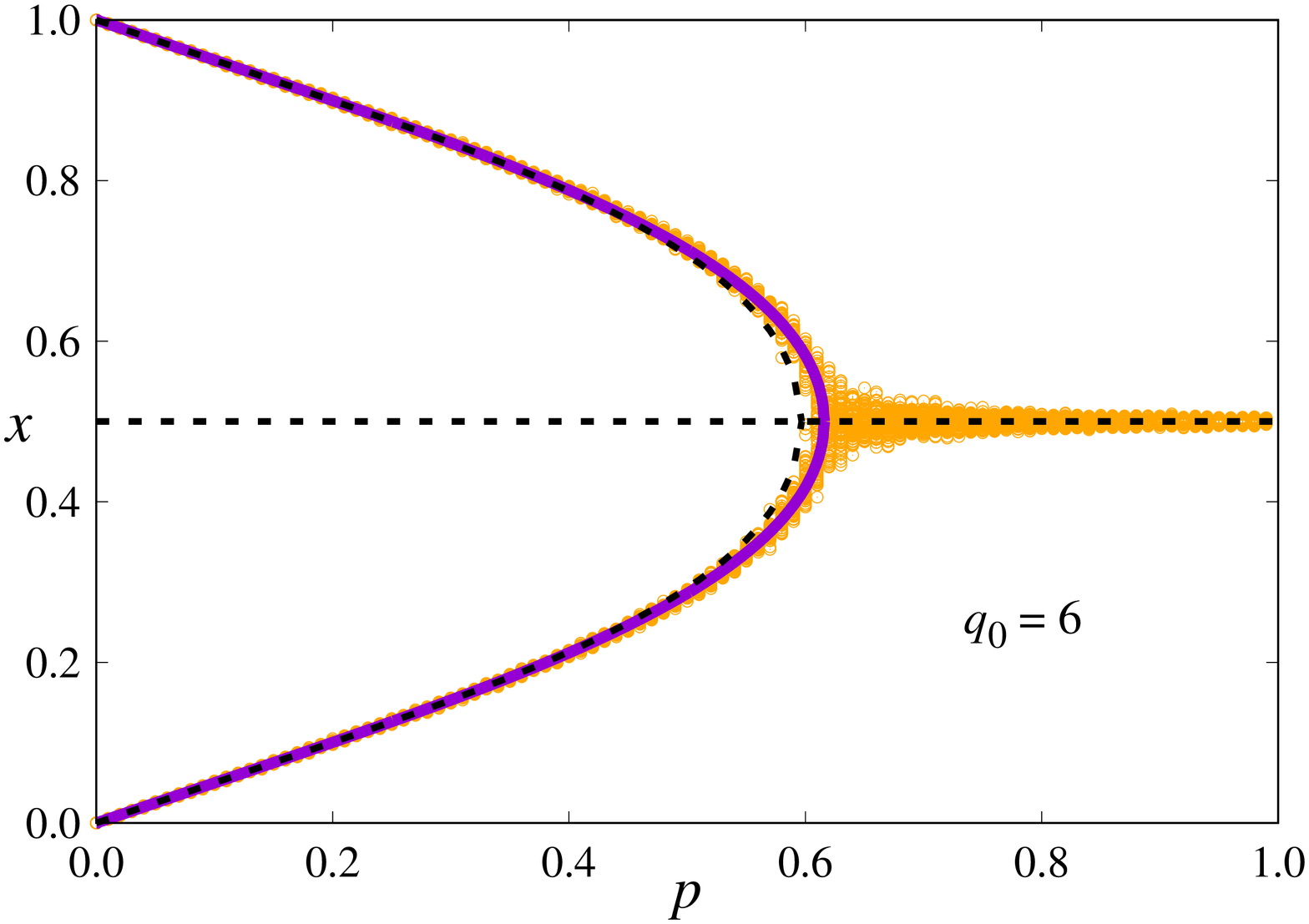}
\end{center}
\caption{{\bf Without repetition.} Steady states $x$ vs. $p$, 
for $\mu=q=12$ and different values of $q_0$. 
Both the pair approximation (lilac solid) and mean-field (black dashed) predictions 
are shown (without distinguishing stable and unstable steady states). 
The (light) orange symbols correspond to numerical simulations on top of 
random regular networks of size $N=10^4$, varying $p$ at intervals $\Delta p =0.01$. 
For each value of $p$, 101 different configurations were chosen, 
each one with fraction of positive nodes $x_0=0.01 j$, with $j=0,1,\ldots, 100$ 
and building a new network. 
For each initial configuration a time average is displayed, 
calculated over the last $10^5$ time units of a trajectory of total length $t=10^7$.
}
\label{fig:12}
\end{figure*}
\begin{figure*}[h!]
\begin{center}
\includegraphics[scale=0.23,angle=-90]{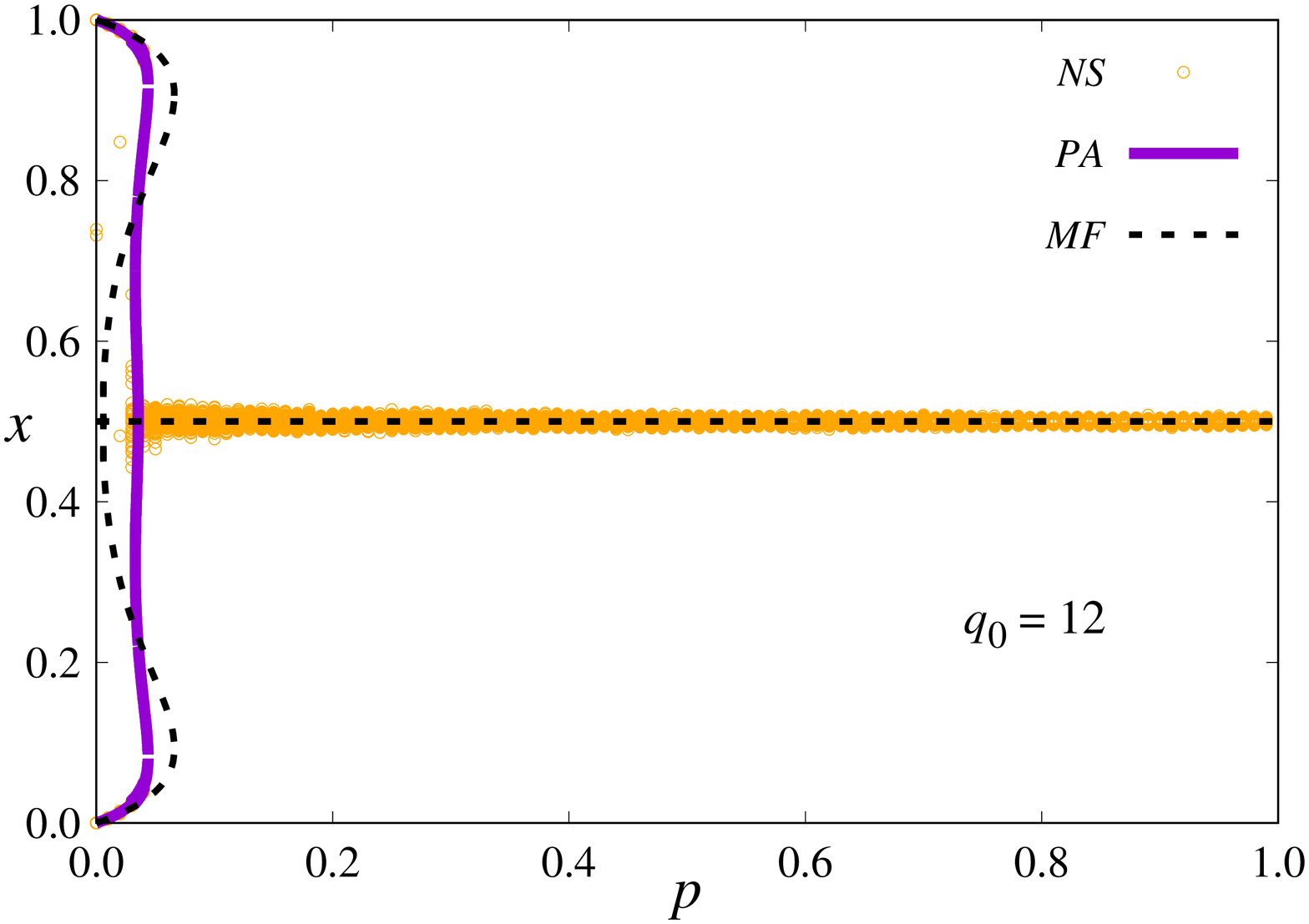}
\includegraphics[scale=0.23,angle=-90]{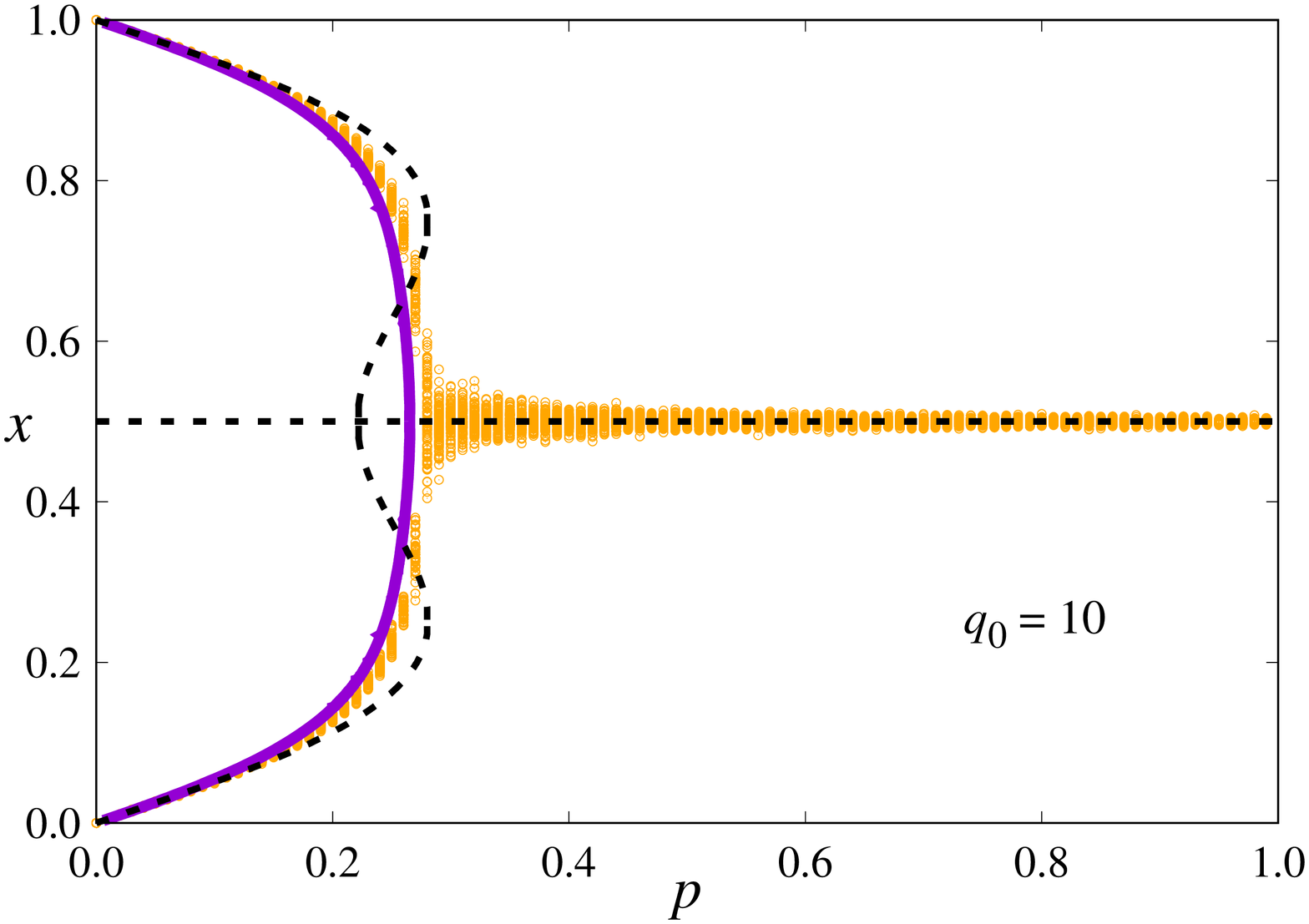}\\
\includegraphics[scale=0.23,angle=-90]{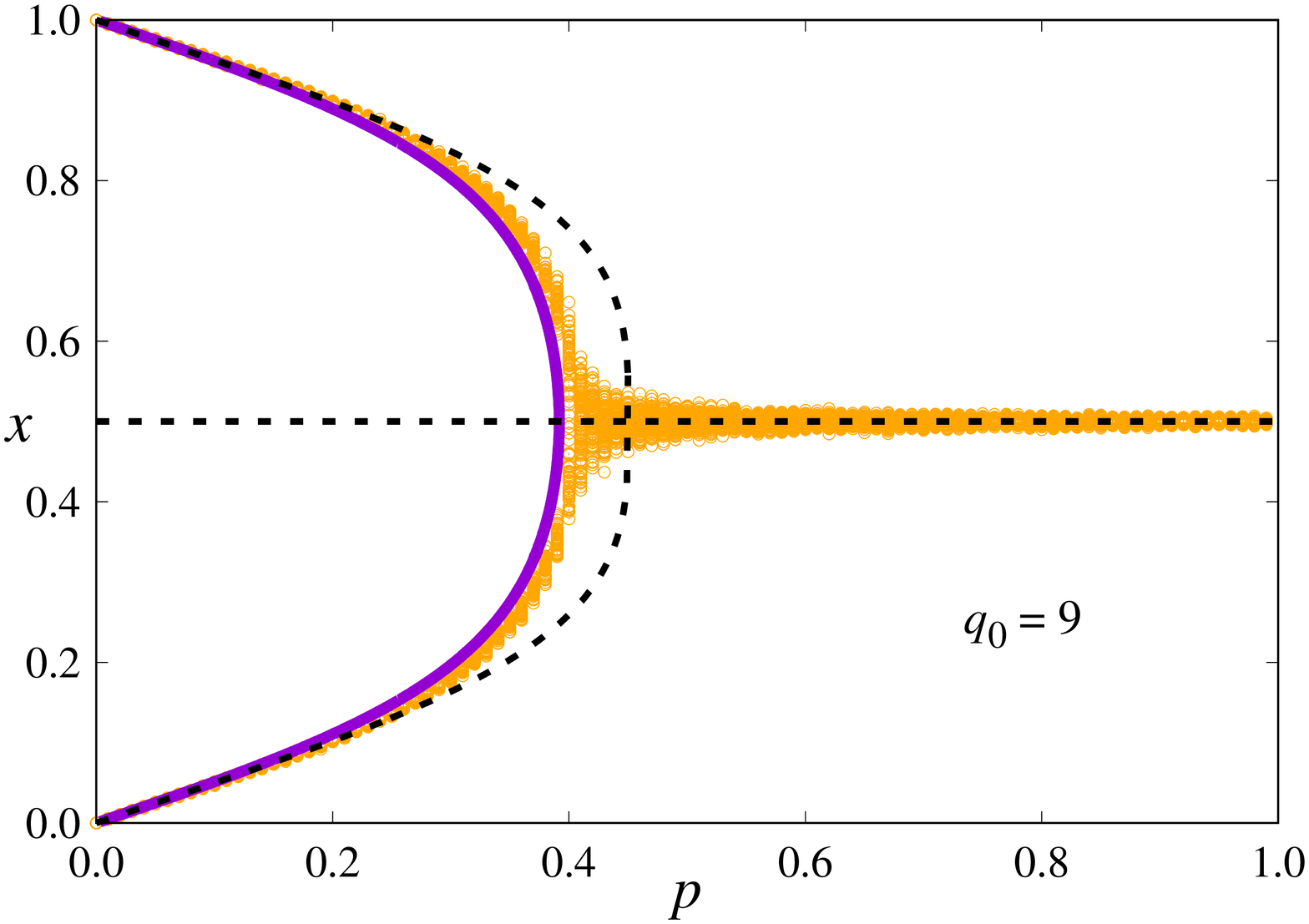}
\includegraphics[scale=0.23,angle=-90]{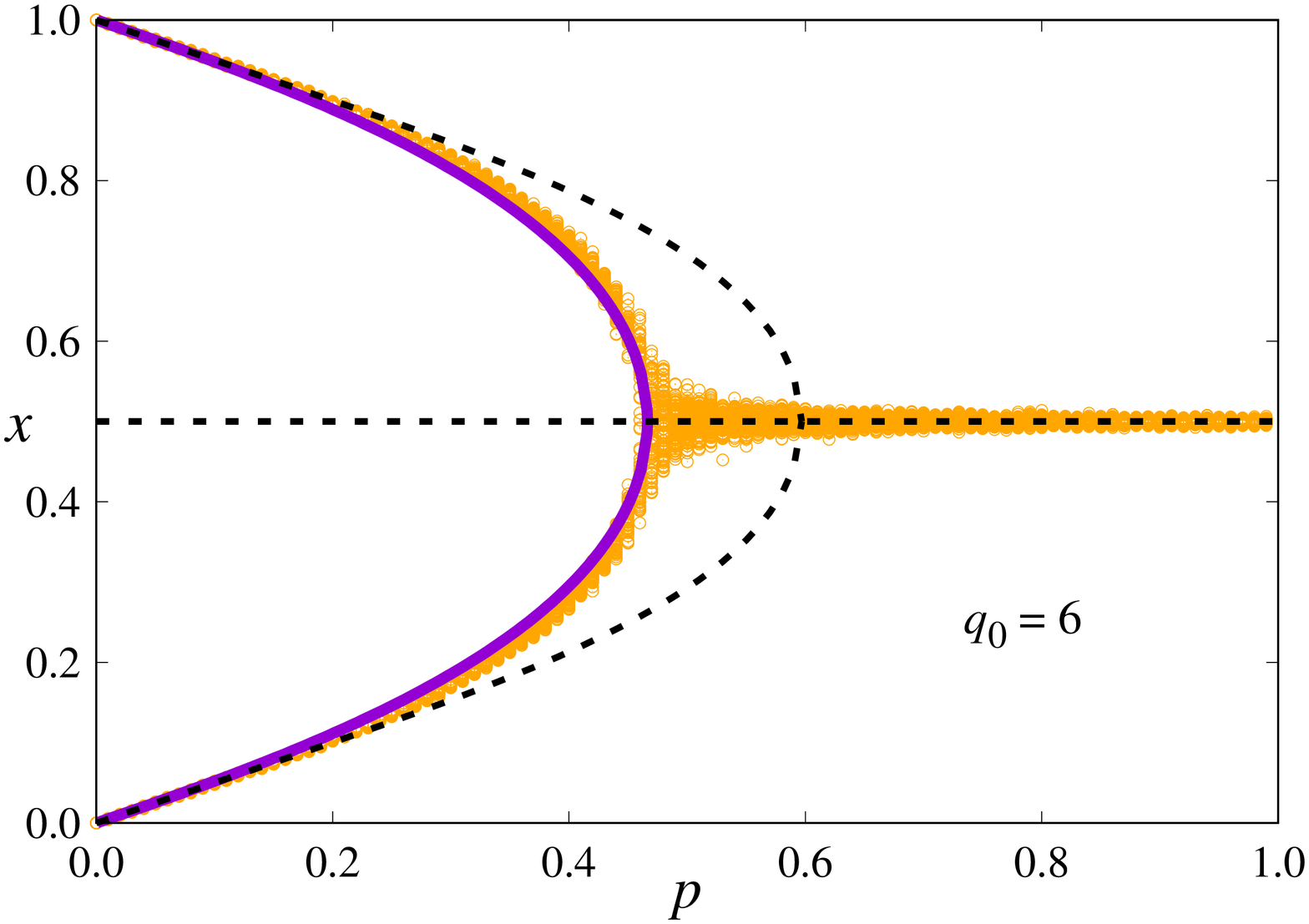}
\end{center}
\caption{{\bf Repetition allowed. } Steady states as in Fig.~\ref{fig:12}. 
}
\label{fig:12rep}
\end{figure*}

In a random regular network each node has the same number of neighbors, $P(k)=\delta(k-\mu)$. It is a suitable network to implement the dynamics based on groups of size $q$, specially in the case without repetition, which imposes constraints in other topologies (i.e. $P(k<q)=0$). Moreover, it is a random network characterized by average clustering coefficient that decays with network size as $C\sim (\mu-1)^2/(N\mu)$), for which the pair approximation is expected to apply well, allowing comparison of simulations with a theoretical framework. The outcomes are illustrated in Figs.~\ref{fig:12} and ~\ref{fig:12rep}, setting $\mu=q=12$, for the versions without and with repetition, respectively.

In connection with Fig.~\ref{fig:12}, which shows the results without repetition, 
the agreement between PA prediction (lilac full line) 
and numerical simulations (light symbols) is very good in all cases, mainly for 
small $q_0\simeq q/2$, while disagreements are more noticeable when $q_0$ increases. 
Note, for instance, for $q_0=10$, 
that the multistability predicted by PA is not observed in numerical simulations. 
In fact, there is a narrow interval around $p\simeq 0.11$, 
with 4 stable steady states in the PA curve (while $x=0.5$ is unstable), 
however, in simulations we observe only tristability 
(the two more ordered PA states and also the disordered one are reached). 
In contrast, for $q_0=9$, the transition is 
also discontinuous, but the state $x=0.5$ is not reached below the 
transition, in agreement with the PA prediction. 
When $q_0=q$ (standard $q$-voter) the system is disordered for any $p$, in accordance with the 
theoretical PA approach, highlighting the crucial role of $q_0$ to promote order.

In the case without ``noise" in the rates $p=0$, we observe a freezing of the dynamics, 
when starting from initial conditions regularly distributed in the 
interval [0,1]. 
The freezing (zero active links) 
occurs at earlier times when $q_0$ approaches $q$, giving rise to 
a spread of points at $p=0$, which become more concentrated when $q_0$ 
decreases.

When repetition is allowed (see Fig.~\ref{fig:12rep}), 
the agreement of PA with numerical simulations is still better 
than in the case without repetition. 
Note that the critical point decreases with respect to 
the MF result when $q_0$ approaches $q/2$, 
in contrast with the case without repetitions where the discrepancy with MF is 
stronger when $q_0$ approaches $q$. The system orders more easily with repetitions, 
except when $q_0\simeq q/2$. 
Also note that discontinuous transitions are less common than in the 
case without repetitions.

A summary of the results in random regular networks is presented in panels B) and C) if Fig.~\ref{fig:diagrams}
that shows the phase diagrams obtained through numerical simulations and PA. 
The critical points were evaluated from the simulations using finite-size scaling analysis 
with precision $\delta p < 0.002$, as illustrated in the Appendix B. 
The MF case is also shown for comparison. 
It is evident that the presence of a network structure has a stronger 
influence in the case with repetition, where the discontinuous transitions are 
less common than in fully connected networks. 
This influence is observed for $\mu=q=6$ and even for $\mu=q=12$ (closer to the MF).
Also note that the PA prediction of critical points is very good, 
which validates our use of analytical expressions to obtain some of the reported results 
(Figs.~\ref{fig:q0effect}-\ref{fig:MFlimit}).

\subsection{Other networks}

We also considered networks such as Erd\H{o}s-R\'enyi, where $P(k)$ 
follows a Poisson distribution $P(k)={\rm e}^{-\mu} \mu^k/k!$, 
and power-law, where $P(k)\sim k^{-a}$.

Recall that in the case without repetition, the pair approximation predicts dependence only on the first moment $\mu$ of the degree distribution. Indeed, in the numerical simulations, we did not detect any significant discrepancy between random regular, Erd\H{o}s-R\'enyi and power-law networks, with the same $\mu$, in agreement with PA predictions. However, we cannot guarantee that it is not due to the large values of $\mu$ used to fulfill the condition $k \ge q$ for all $k$. 
In order to go far away from the MF limit where the effects of repetitions and other features related to structure vanish, 
we must decrease $\mu/q$. But, then, the probability of finding nodes with degree $k<q$ increases. 
A lower bound $k_{min} \ge q$ can be imposed, but 
this restricts the minimal value of $\mu$ in Erd\H{o}s-R\'enyi networks or the 
shape of $P(k)$, like heavy-tailed ones, whose exponent is approximately given by
$a \simeq (2\mu-k_{min})/(\mu - k_{min})$. 
Therefore, the accessible values of the parameters may remain limited to a region where 
the effects of the degree distribution beyond the first moment are not significant. 
This is the case of the results reported by J\c{e}drzejewski~\cite{qvoterPA} 
about the standard $q$-voter model without repetitions, 
keeping $\mu$ much larger than $q$. 
Then no significant differences were reported for random regular, Erd\H{o}s-R\'enyi or even scale-free networks. 
Only for Watts-Strogatz~\cite{WS} networks with low rewiring probability, 
hence higher clustering, differences were observed as expected. 

\begin{figure*}[h!]
\begin{center}
\includegraphics[scale=0.23,angle=-90]{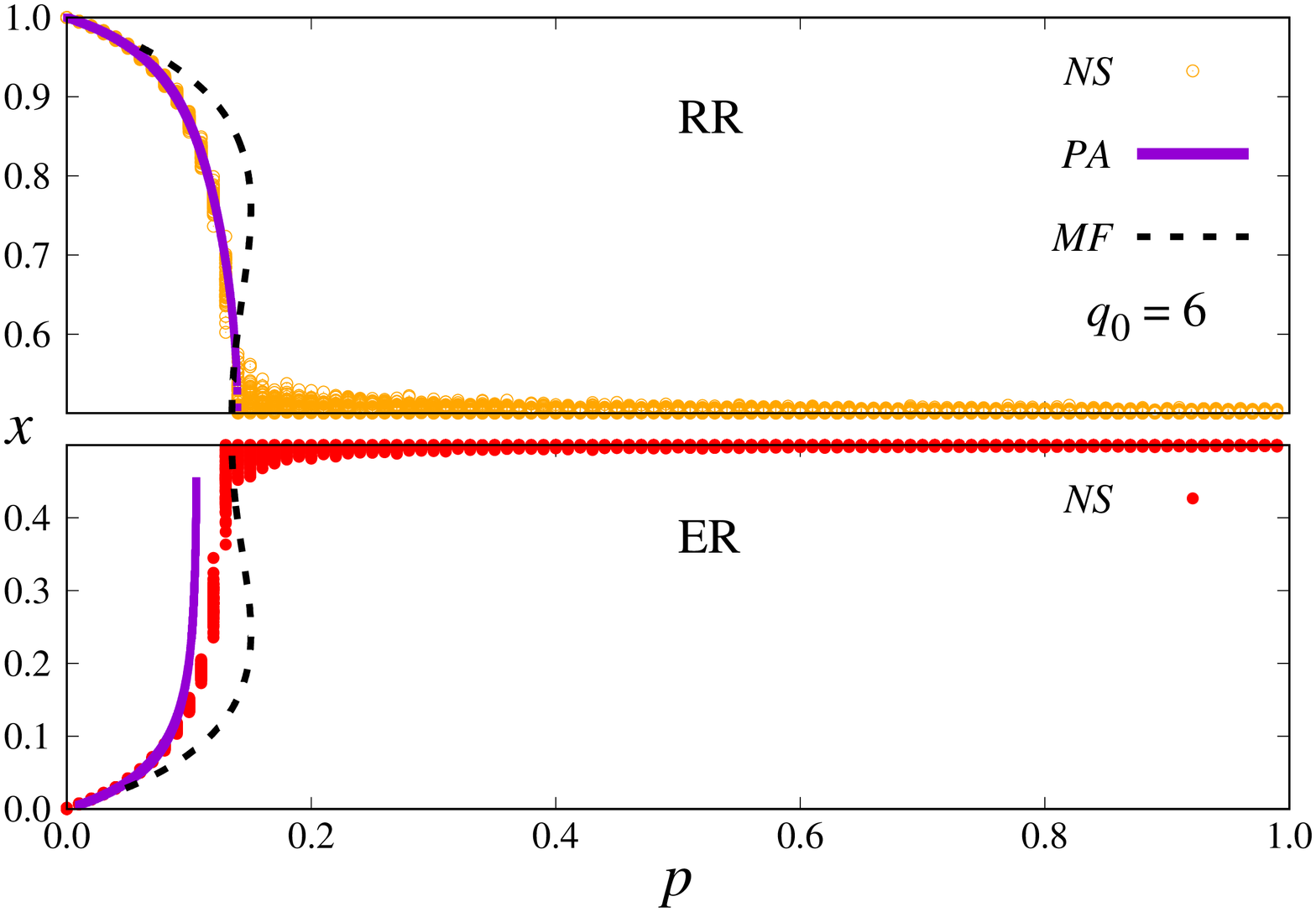} 
\includegraphics[scale=0.23,angle=-90]{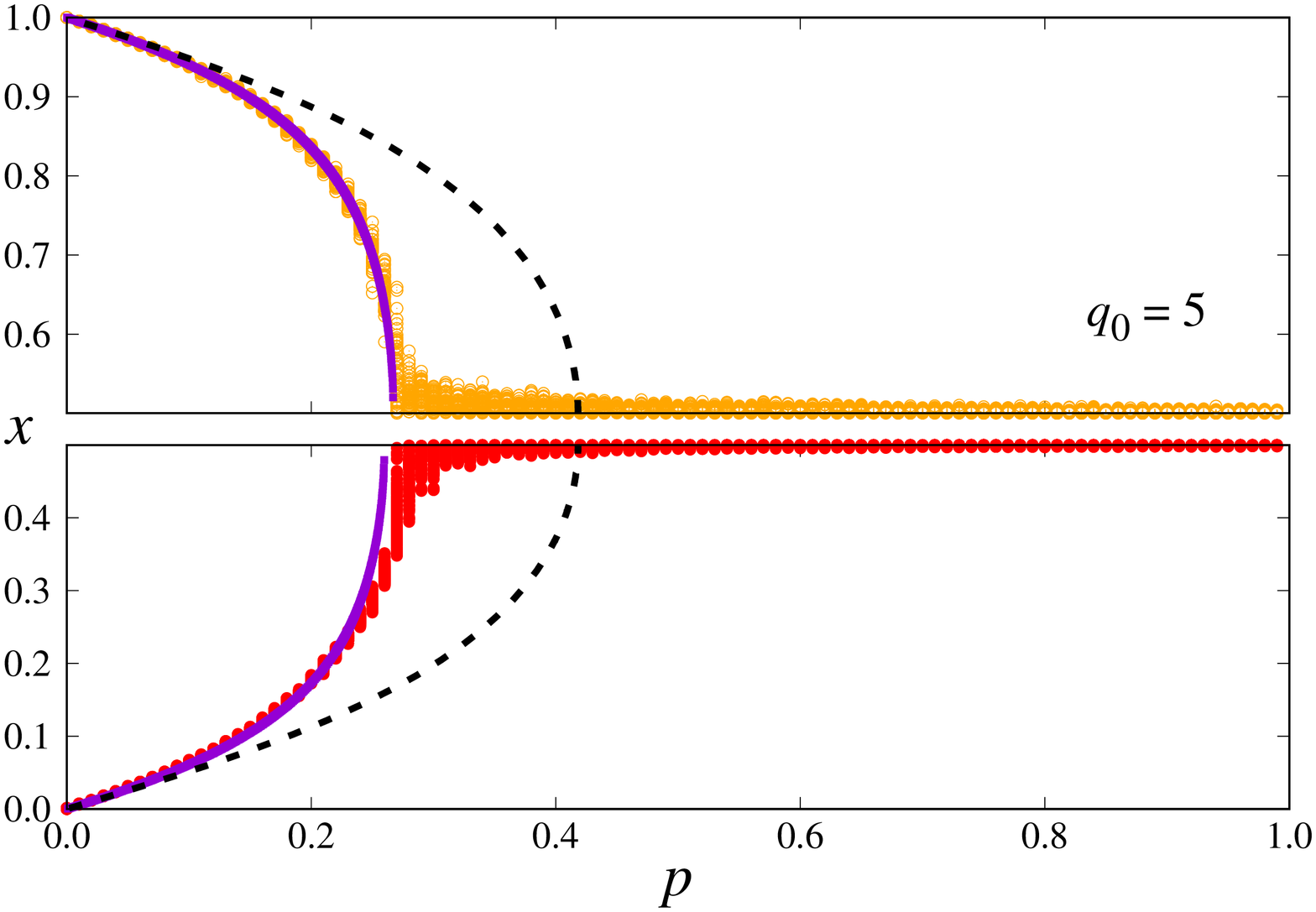} \\
\includegraphics[scale=0.23,angle=-90]{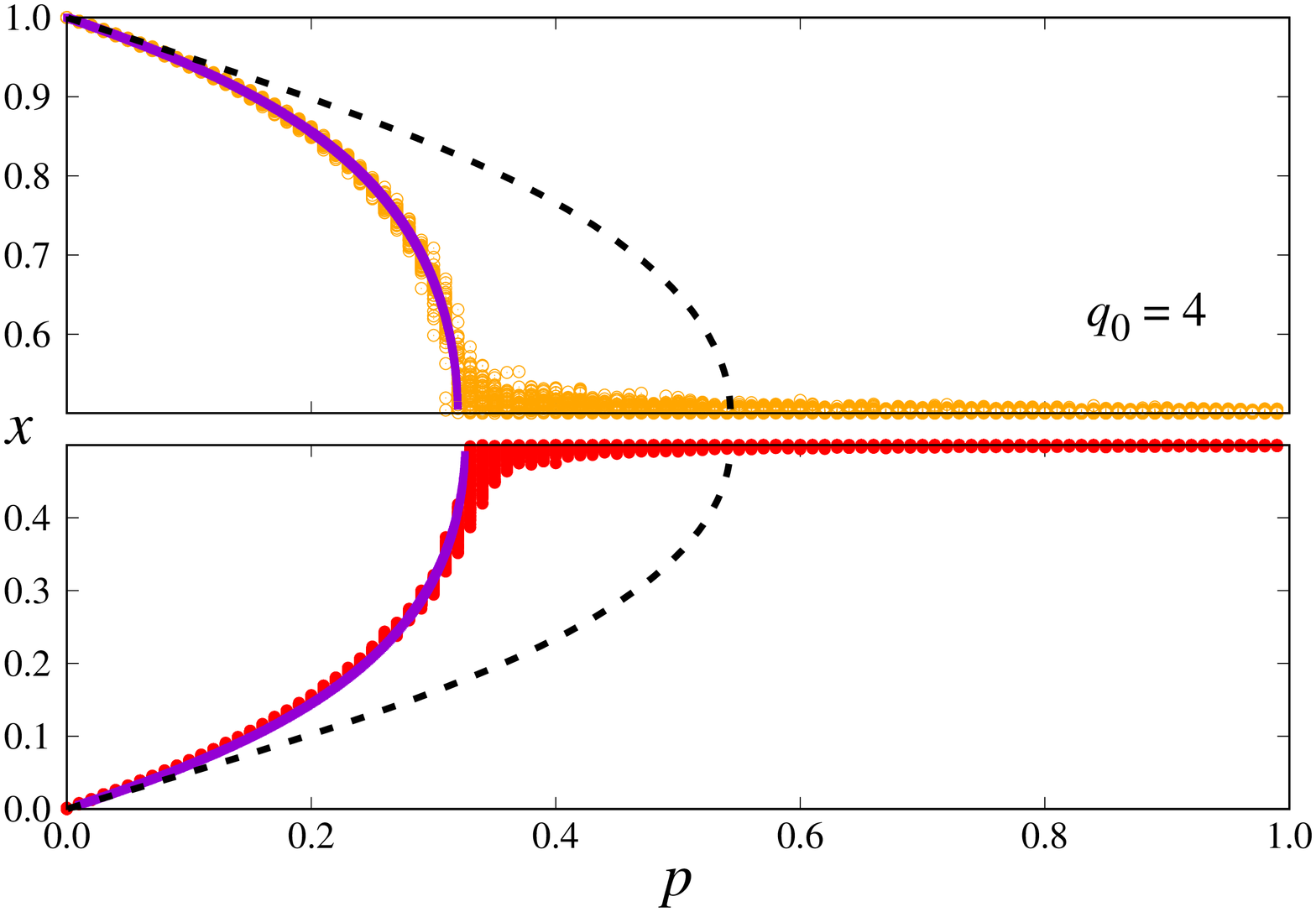} 
\includegraphics[scale=0.23,angle=-90]{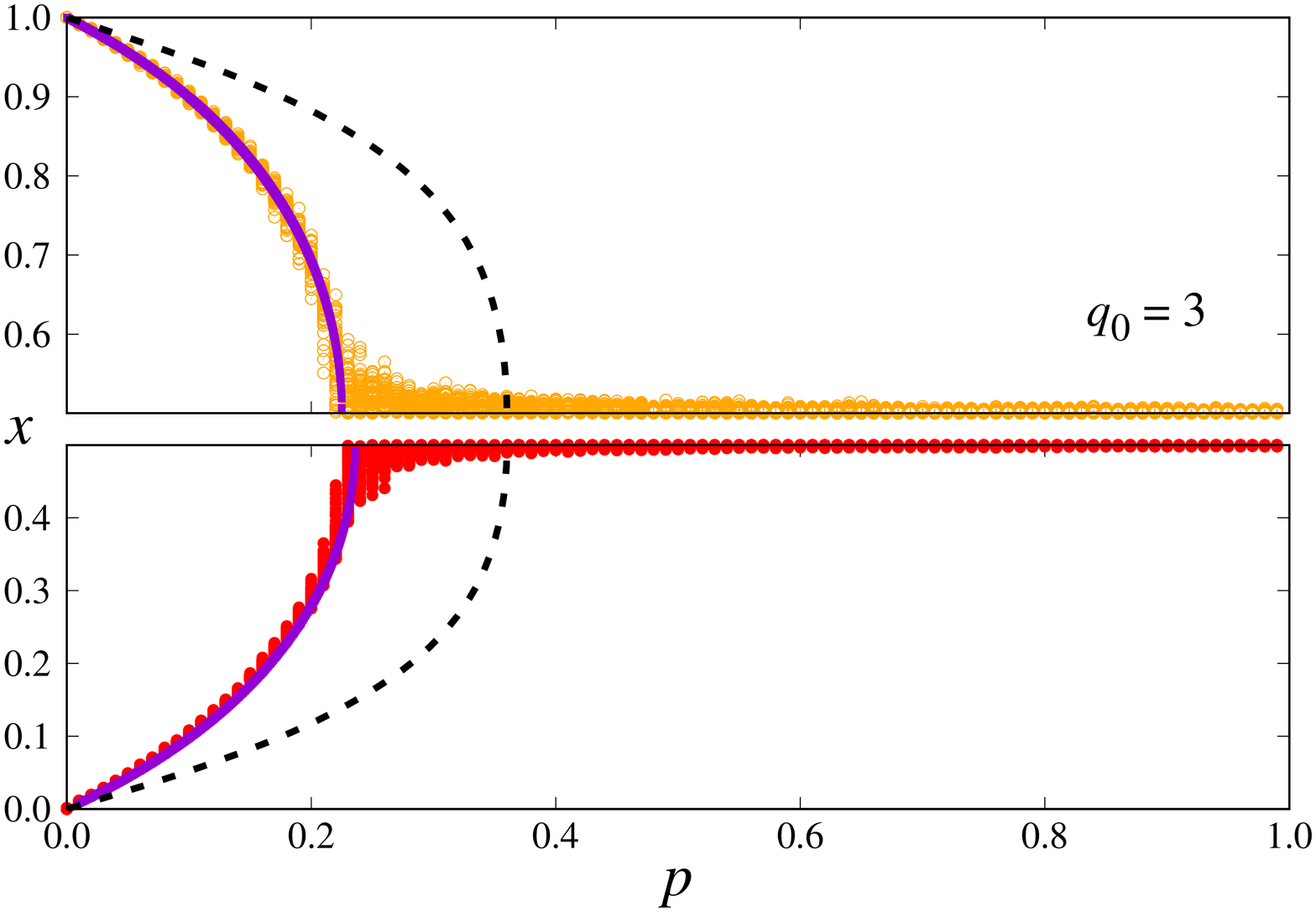} 
\end{center}
\caption{Stability diagrams $x$ vs. $p$, for $q=6$, and different values of $q_0$, 
with $\mu=q$, with repetition. 
Simulations and calculations are for random regular (upper half) and Erd\H{o}s-R\'enyi 
(lower half of each panel) networks. 
Other details are as in Fig.~\ref{fig:12}.
}
\label{fig:RRER}
\end{figure*}
\begin{figure*}[h!]
\begin{center}
\includegraphics[scale=0.23,angle=-90]{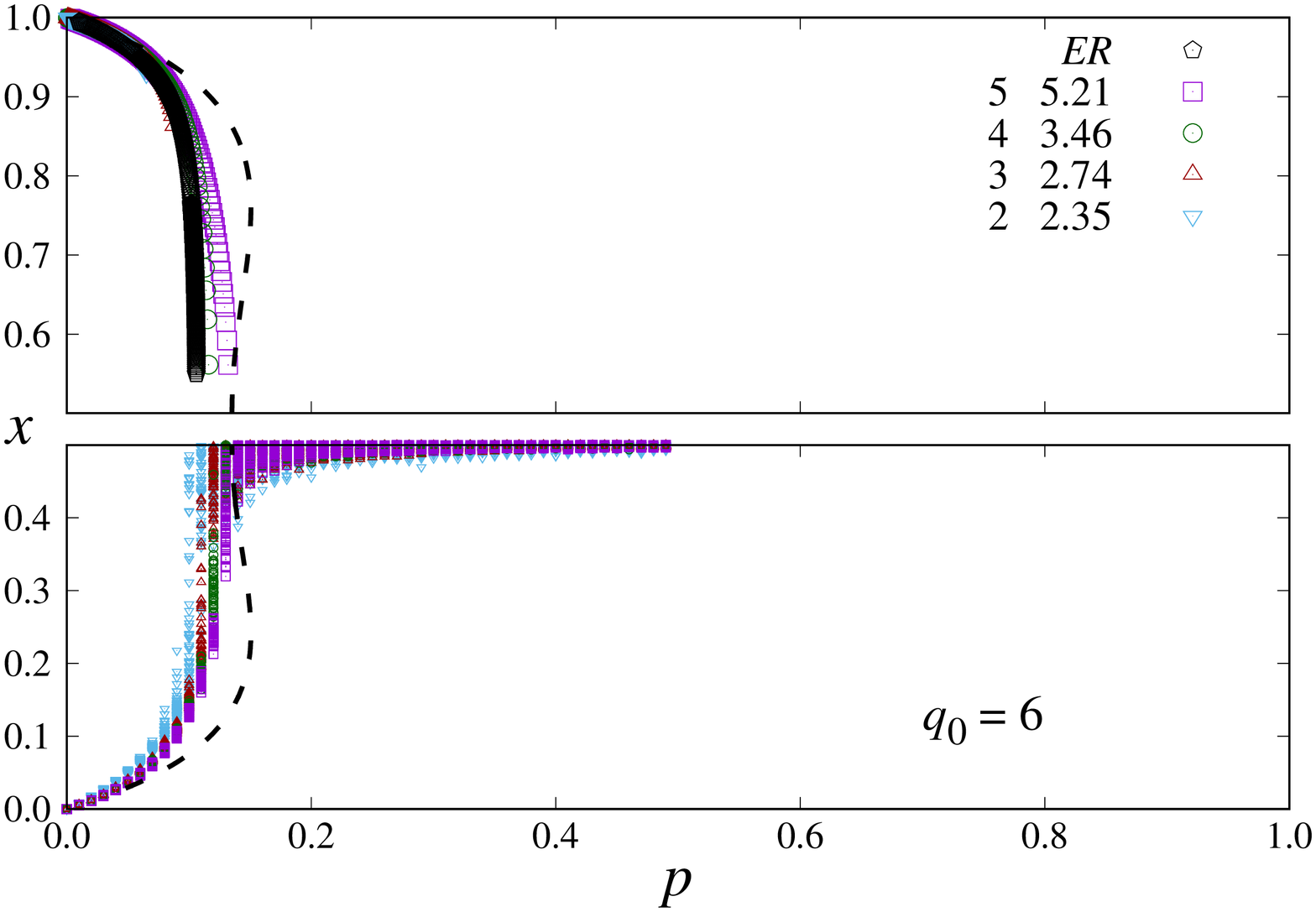} 
\includegraphics[scale=0.23,angle=-90]{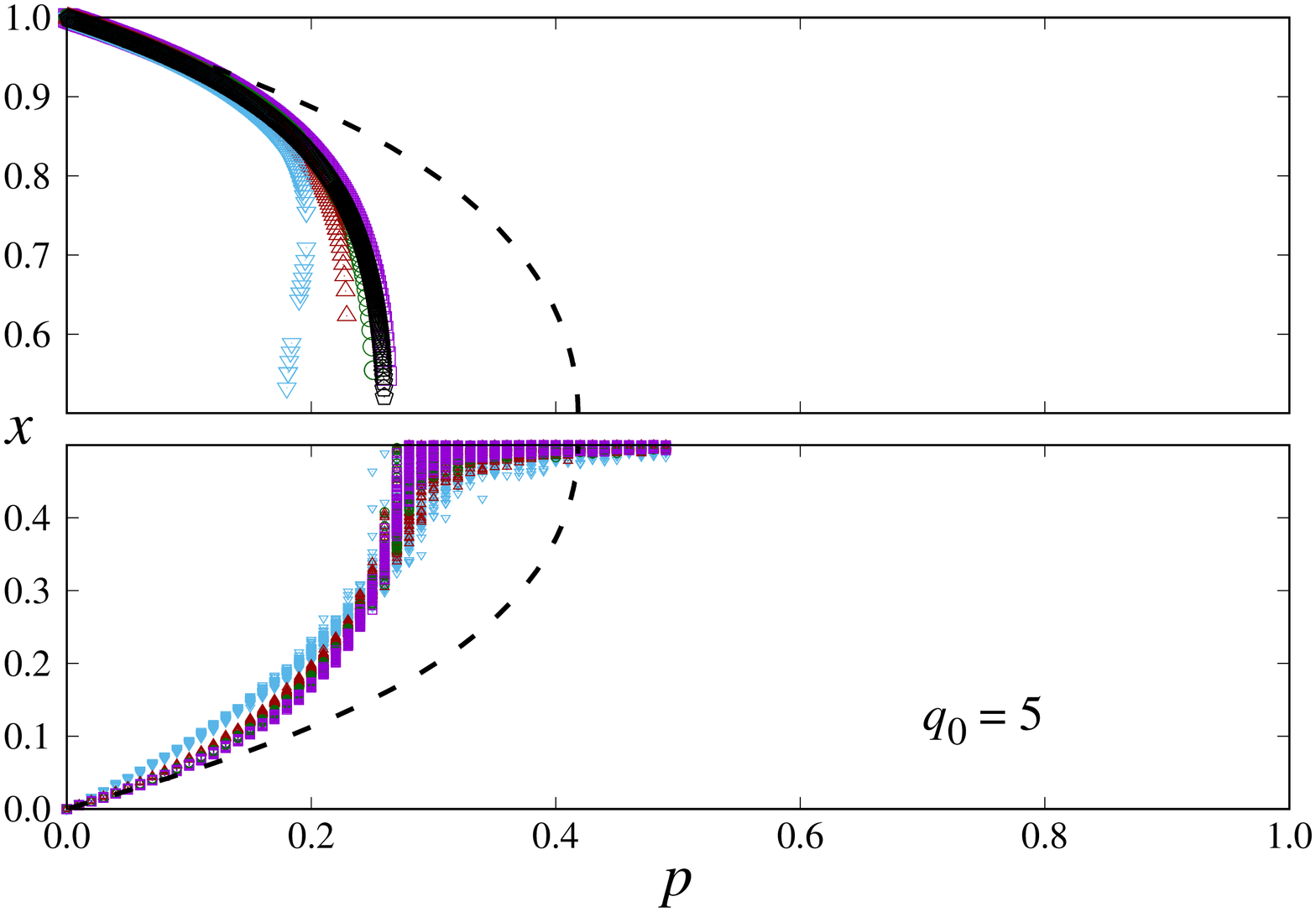} \\
\includegraphics[scale=0.23,angle=-90]{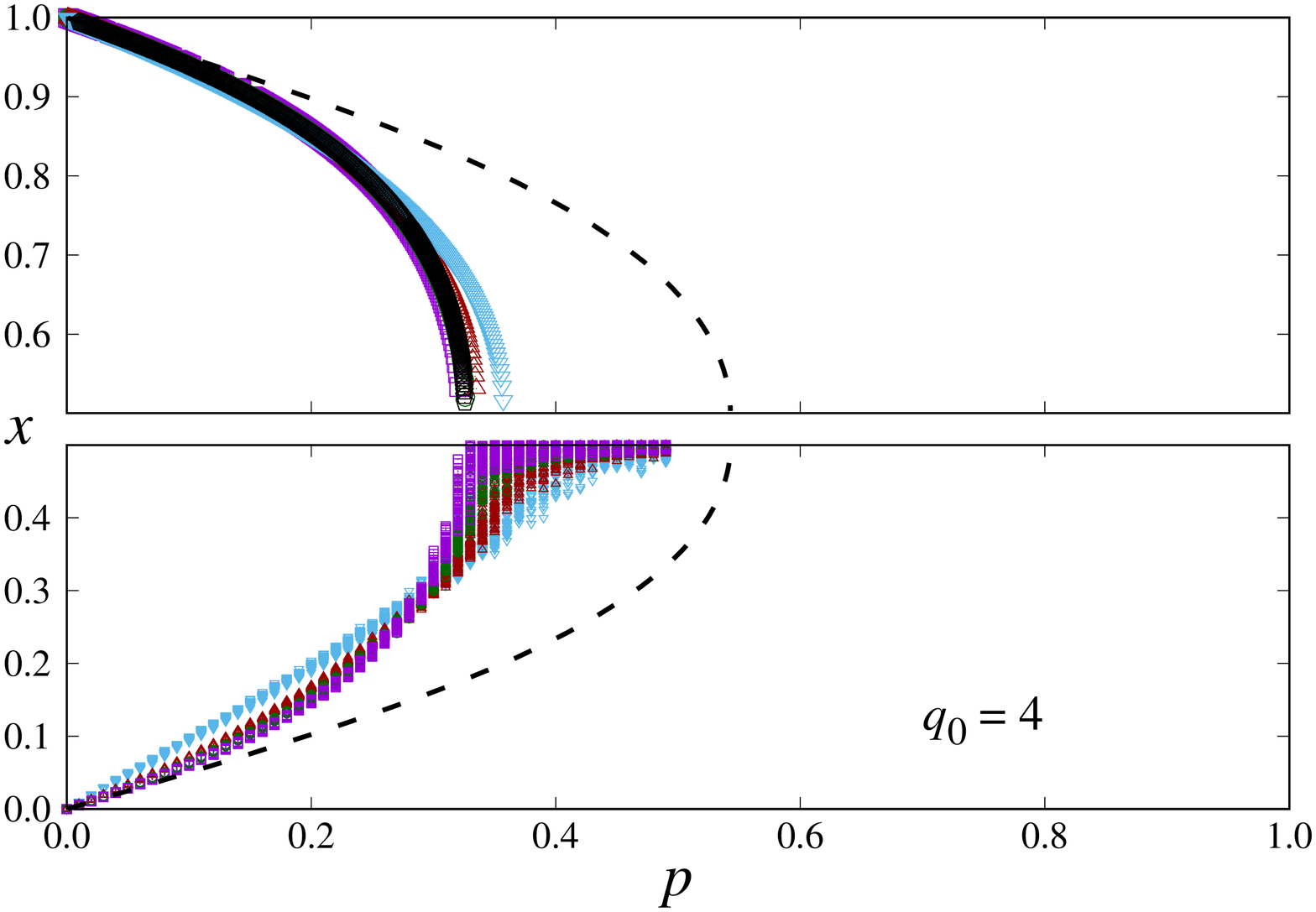} 
\includegraphics[scale=0.23,angle=-90]{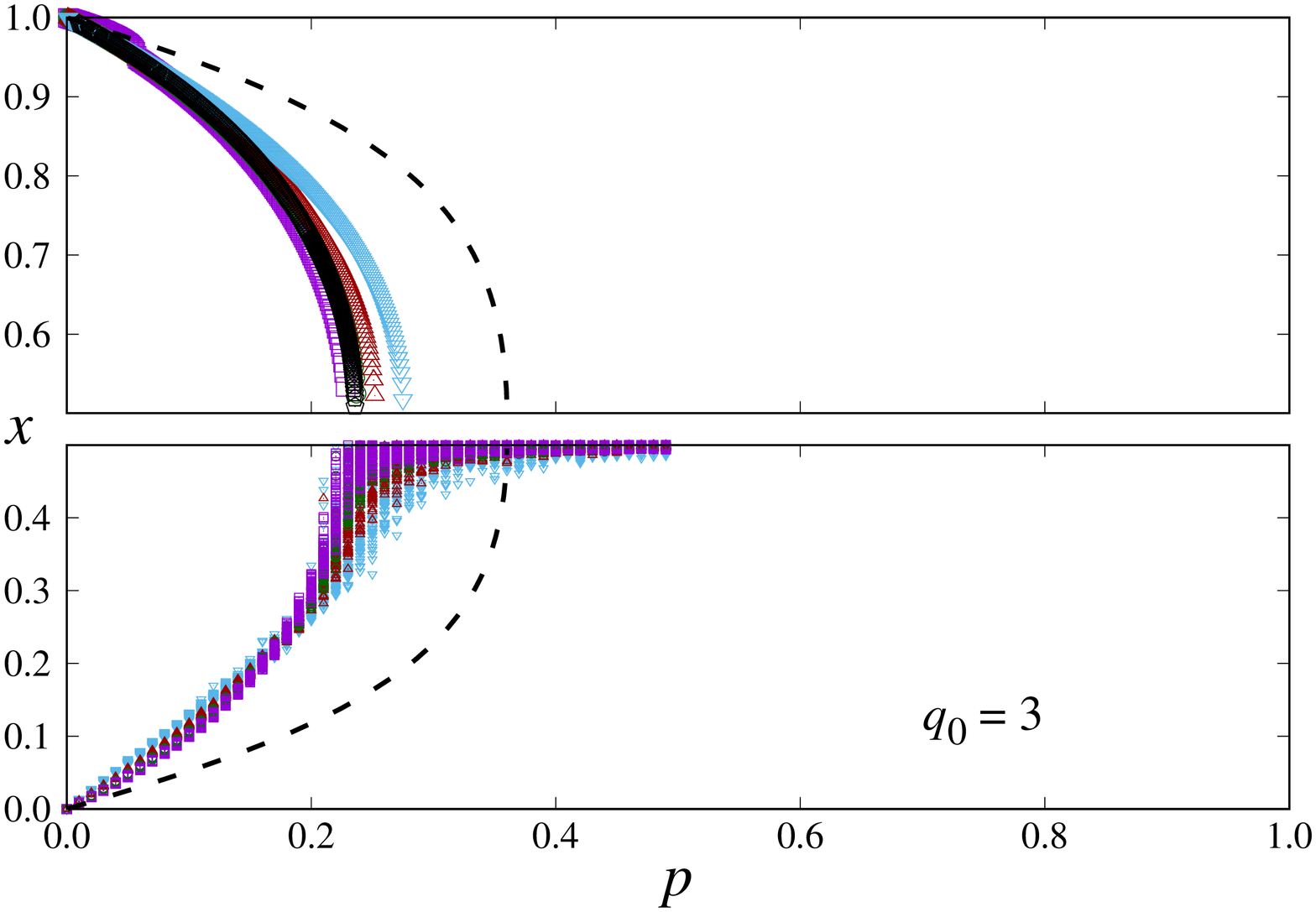} 
\end{center}
\caption{Stability diagrams $x$ vs. $p$, for $q=6$, and different values of $q_0$, 
with $\mu=q=6$, with repetition. 
PA calculations (upper half) are for power-law degree distributions 
$P(k) \propto 1/k^a$, for $k_{min} \le k \le k_{max}=N/5$, and zero otherwise. 
In the legend, we show the values of $k_{min}$ and the corresponding value of $a$ (truncated at the 
second decimal place), such that $\mu=6$. Results on 
Erd\H{o}s-R\'enyi networks and mean-field results are also shown for comparison. 
Other details are the same as in previous figures. 
Numerical-simulation results for $N=10^4$ and $\langle k\rangle=6$ 
are shown in the lower half of each graph. 
}
\label{fig:PL}
\end{figure*}

But such limitations are absent if repetition is allowed. Moreover, in such case, PA predicts a dependency on further details 
of the network structure beyond $\mu$, which makes more motivating the 
study of the dynamics in different topologies. 
Then analytical calculations and simulations for Erd\H{o}s-R\'enyi and 
power-law networks, with repetition, were performed. 

As shown in Fig.~\ref{fig:RRER}, small, but visible, discrepancies between 
Erd\H{o}s-R\'enyi and random regular results exist for $q=\mu=6$. 
These differences become negligible if we increase the average degree, for example, to $\mu=3q=18$.

In Erd\H{o}s-R\'enyi networks, whose average clustering behaves as $C\sim \mu/N$, 
there is still a good agreement 
between simulations and the corresponding analytical results of the pair approximation. 
Note that a change of the nature of the transition, 
from discontinuous (in MF, equivalent to $\mu\to \infty$) to continuous, 
occurs for $q=q_0=6$, in both networks. 
The same change is observed in random regular networks for $q=12$ and $q_0=10$ 
(see Figs.~\ref{fig:MFlimit} and \ref{fig:12}).
The critical points depart more from the MF value at $q_0=4$, and are always 
smaller than in MF, consistent with the intuition that the system 
has more difficulty in ordering when networks have structure, compared to 
the MF scenario. 
Note also that the deviation between the critical points 
for random regular and Erd\H{o}s-R\'enyi is not systematic but changes sign, between $q_0=3$ and $6$.

The results for random regular and Erd\H{o}s-R\'enyi are very close, even for $\mu=6$, 
and become even closer when $\mu$ increases (because the MF is approached). 
In fact, the influence of the negative moments becomes irrelevant when the network  
loses structure. 
We expect that the impact of these negative moments increases for power-law decaying $P(k)$. 
Then we considered $P(k) ={\cal N}/k^a$, for $k_{min}\le k \le k_{max}$, and null otherwise, 
where ${\cal N}$ is a normalization factor. 
For given values of the minimal degree $k_{min}$, we adjusted $a$ 
to obtain the desired average degree $\mu$. 
Results are shown in Fig.~\ref{fig:PL}.
For $k_{min}=5$, the power-law exponent is large and the results resemble more those of 
random regular and Erd\H{o}s-R\'enyi networks. As $k_{min}$ decreases (hence exponent $a$ decreases), 
we observe that in all cases the curves separate from random regular 
but two distinct behaviors are observed. While for $q_0=3$ and 4, the critical 
point shifts towards the MF limit, 
for $q_0=5$ and 6, contrarily the critical values decrease, becoming $p_c=0$ at the $q$-voter 
limit $q_0=q=6$. This may be due to the 
increasing presence of nodes with low connectivity concomitantly with the fattest tail. 
Nodes with low connectivity have influence only on few nodes, hampering order, 
while for highly connected nodes, 
the requirement of local consensus for change (when $q_0\simeq q$) 
is more difficult to be attained than for the simple majority (when $q_0\simeq q/2$).

\section{Final remarks}
\label{sec:remarks}

We investigated the effect of network topology on the dynamics of the 
threshold $q$-voter model with level of independence $p$. 
The introduction of structure makes relevant the study of the possibility 
of repetitions in the selection of $q_0$ amongst $q$ opinions. 
Then, we considered two implementations, either allowing repetition or not.

Analytical results were derived using the PA, for random networks with arbitrary 
degree distribution $P(k)$.
Simulations of the threshold $q$-voter dynamics were obtained mainly on random regular networks, 
which is more adequate to implement the group dynamics. 
But, simulations on Erd\H{o}s-R\'enyi and power-law networks were also performed. 

The structure of random-regular networks has a stronger 
influence in the case with repetition, where the discontinuous transitions are 
less common than in fully connected networks (see Fig.~\ref{fig:diagrams}). 
The threshold parameter $q_0$ has a crucial interplay with network structure and repetition effects. 
Without repetition, 
deviations from MF behavior are more pronounced when $q_0\simeq q$. 
Differently, with repetition, the maximal deviation from MF occurs for $q_0\simeq q/2$. 
(See Figs.~\ref{fig:diagrams} and \ref{fig:MFlimit}.) 
Furthermore, depending on the values of $q,q_0$, 
while transitions which are continuous in the MF can become discontinuous 
in networks without repetition, the opposite behavior can be observed with repetition 
(see Fig.~\ref{fig:MFlimit}).
As can be seen in the comparative phase diagrams of Fig.~\ref{fig:diagrams}, and also in 
the bifurcation diagrams of Fig.~\ref{fig:12rep}, 
the critical points estimated through the PA 
are in very good agreement with those obtained through finite-size scaling analysis 
of the outcomes of simulations in random regular networks. 
This is specially true when $\mu$ increases 
approaching the exact MF result where repetition and other issues related to 
structure become irrelevant. 
Forbidding repetition, PA works better when $q_0$ approaches $q/2$, 
and spurious results, such as multistability beyond 3 states 
are observed in cases with $q_0 \simeq q$ and $\mu\simeq q$. 
Larger deviations are expected in networks with higher correlations.

For Erd\H{o}s-R\'enyi and power-law networks, 
there is also a good agreement between the simulations and the predictions of the pair approximation. 
When repetitions are forbidden, we observed agreement of the results of simulations 
for (Erd\H{o}s-R\'enyi and power-law) networks with same values of $\mu$ as predicted by the 
PA. However, the structures that can be visited are limited by the constraint $k\ge q$. 
This is in agreement with results reported for the $q$-voter~\cite{qvoterPA}, 
where discrepancies were observed only for networks with large clustering coefficient.
Allowing repetitions, the PA predicts dependency 
on network structure beyond the average degree $\mu$, in accord with simulations. 
These effects are weak in Erd\H{o}s-R\'enyi networks (see Fig.~\ref{fig:RRER}) but stronger in networks 
with power-law degree distribution (see Fig.~\ref{fig:PL}), where long tails are concomitant 
with high probability of poorly connected nodes to realize small values of $\mu$, far away from the MF. 
Also in this case $q_0/q$ plays a crucial role. While for $q_0\simeq q/2$, 
by decreasing the power-law exponent, 
the critical point increases towards the MF value, 
for $q_0 \simeq q$, the opposite effect occurs, and the 
appearance of highly connected nodes promotes disorder.

\section*{Acknowledgements}
Partial financial support has been received from the Agencia Estatal de Investigaci\'on (AEI, Spain) and Fondo Europeo de Desarrollo Regional (FEDER, UE), under Project PACSS (RTI2018-093732-B-C21/C22) and the Maria de Maeztu Program for units of Excellence in R\&D (MDM-2017-0711). A.F.P. acknowledges support by the Formaci\'on de Profesorado Universitario (FPU14/00554) program of Ministerio de Educaci\'on, Cultura y Deportes (MECD) (Spain). 
A.R.V. and C.A.  received partial financial support from Conselho Nacional de Desenvolvimento Cient\'{\i}fico e Tecnol\'ogico (CNPq). C.A. also acknowledge partial financial support 
from Coordena\c c\~ao de Aperfei\c coamento de Pessoal de N\'{\i}vel Superior (CAPES), 
and Funda\c c\~ao de Amparo \`a Pesquisa do Estado do Rio de Janeiro (FAPERJ).

\appendix

\begin{widetext}
\section*{Appendix A. Some mathematical expressions}
\label{appB}
\setcounter{equation}{0}
\renewcommand{\theequation}{A.\arabic{equation}}
We start from the sum:
\begin{equation}\label{app_suma1}
\sum_{j=q_0}^q {q\choose j}a^jb^{q-j}={q\choose q_0}a^{q_0}b^{q-q_0}{_2F_1}\left(1,q_0-q;q_0+1;\frac{-a}{b}\right).
\end{equation}
From which it follows
\begin{eqnarray}
\sum_{j=q_0}^q {q\choose j}x^j(1-x)^{q-j}=g_1(x;q,q_0),
\end{eqnarray}
where we have defined the family of functions
\begin{eqnarray}\label{eq:gs}
g_s(x;q,q_0)={q\choose q_0}x^{q_0}(1-x)^{q-q_0}{_2F_1}\left(s,q_0-q;q_0+1;\frac{-x}{1-x}\right).
\end{eqnarray}
Taking the derivative of Eq. (\ref{app_suma1}) with respect to $a$ we obtain:
\begin{equation}\label{app_suma2}
\sum_{j=q_0}^q {q\choose j}ja^jb^{q-j}=a\frac{d}{da}\sum_{j=q_0}^q {q\choose j}a^jb^{q-j}.
\end{equation}
Using Eq. (\ref{app_suma1}), the known properties of the derivative of the hypergeometric function, and replacing $a\to x$, $b\to1-x$ we arrive after some algebra at:
\begin{eqnarray}
\sum_{j=q_0}^q {q\choose j}jx^j(1-x)^{q-j}=q_0g_1(x;q,q_0)+g_2(x;q,1+q_0).
\end{eqnarray}

In order to perform the averages $\langle F(\ell;k,q,q_0,p)\rangle_{\rho_i}$ and $\langle \ell F(\ell;k,q,q_0,p)\rangle_{\rho_i}$ appearing in 
Eqs. (\ref{eq:rho},\ref{eq:x}) after replacing the probabilities Eq. (\ref{eq:f}) using the binomial distribution ${k\choose \ell}\rho_i^l(1-\rho_i)^{k-\ell}$, we use the following results:
\begin{eqnarray}
\sum_{\ell=0}^k{k-q\choose \ell-j}\rho^{\ell}(1-\rho)^{k-\ell}=\rho^j(1-\rho)^{q-j},
\end{eqnarray}
\begin{eqnarray}
\sum_{\ell=0}^k\ell{k-q\choose \ell-j}\rho^{\ell}(1-\rho)^{k-\ell}=\rho^j(1-\rho)^{q-j}\left[j+(k-q)\rho\right].
\end{eqnarray}
Replacement of these results in Eq. (\ref{eq:x}) and the definition of the average $\mu=\sum_k kP(k)$ leads straightforwardly to Eq. (\ref{eq:without}) where
\begin{align}
G_2(z;q,q_0,\mu)\equiv\left[\mu-2q_0-2(\mu-q)z\right]g_1(z;q,q_0)+g_2(z;q,1+q_0).
\end{align}
\end{widetext}

\section*{Appendix B. Evaluation of critical points from numerical simulations}
\label{app}
\setcounter{equation}{0}
\renewcommand{\theequation}{B.\arabic{equation}}

\begin{figure}[h!]
\begin{center}
\includegraphics[scale=0.3,angle=-90]{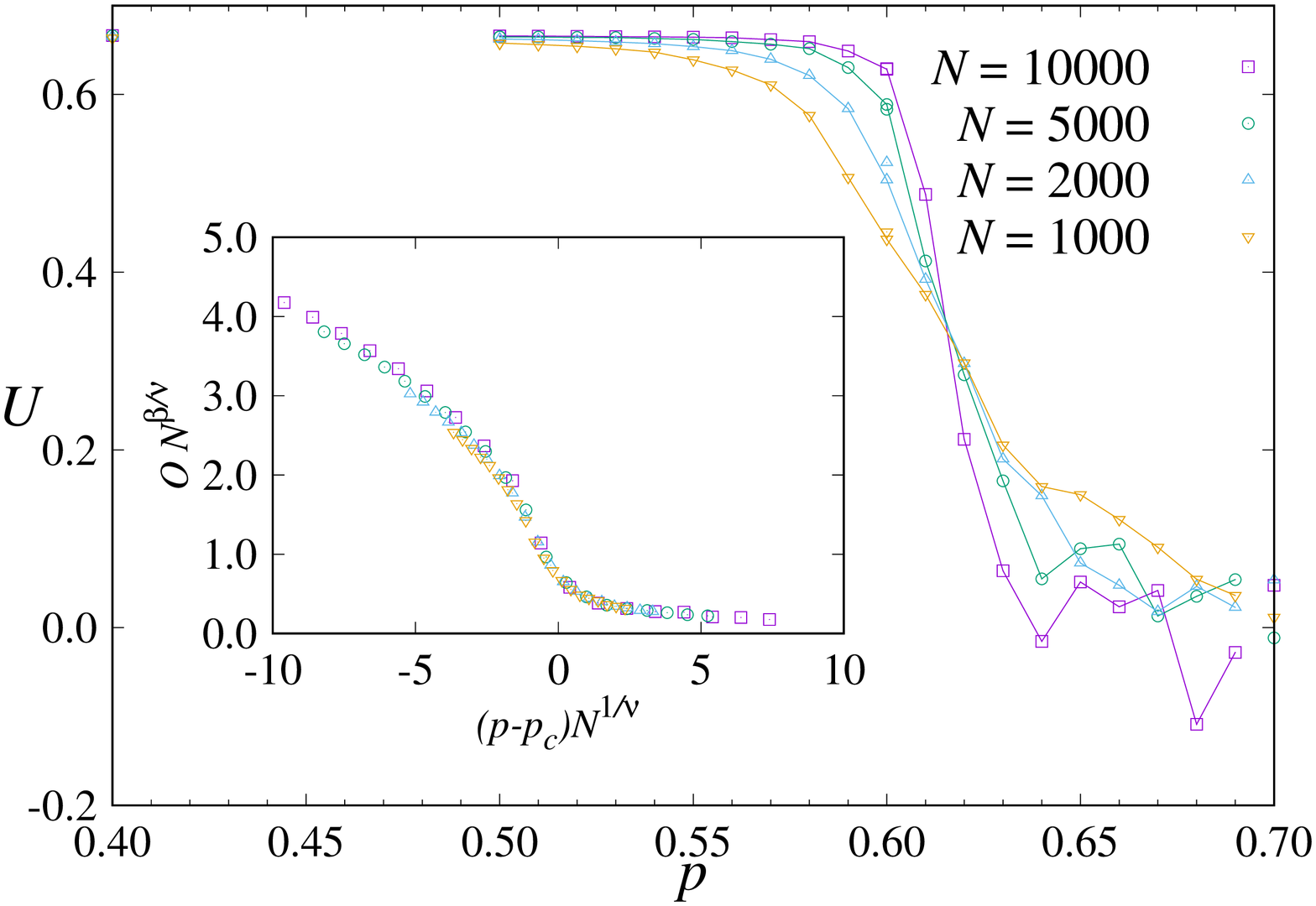} \\
\includegraphics[scale=0.3,angle=-90]{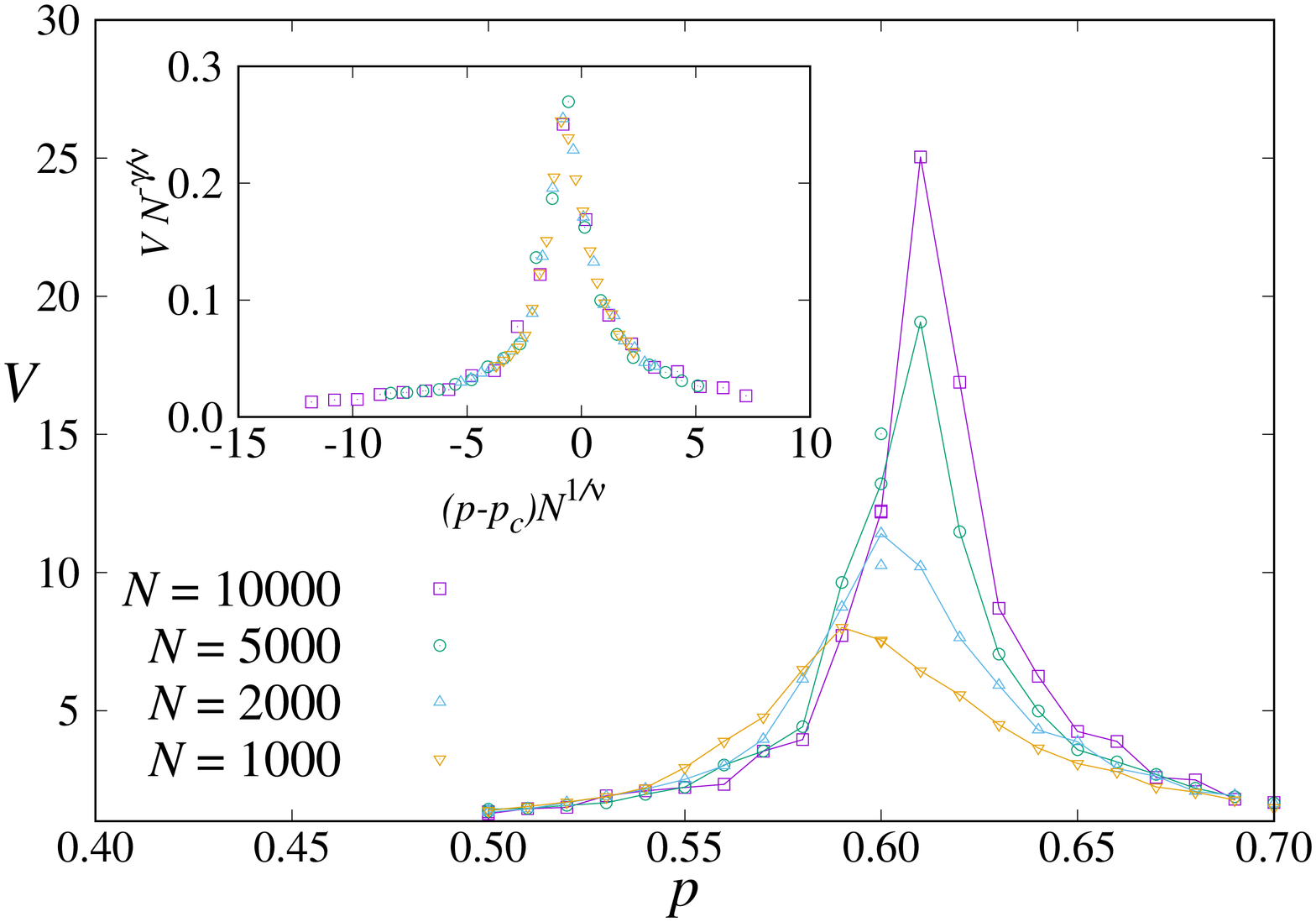} 
\end{center}
\caption{Continuous transition in random regular networks for $q=12$ and $q_0=7$, without repetition. 
Binder cumulant, scaled order parameter, and susceptibility vs. noise $p$ 
for the network sizes indicated on the figure. 
For each value of $p$, and each size, 100 configurations were chosen with $x(0)=0.5$.
We obtained $\beta=1/2$, $\gamma=1$, $\nu=2$~\cite{biswas12,nuno_celia_keom}, and $p_c = 0.616$. 
}
\label{fig:fss-continuous}
\end{figure}

\begin{figure}[h!]
\begin{center}
\includegraphics[scale=0.3,angle=-90]{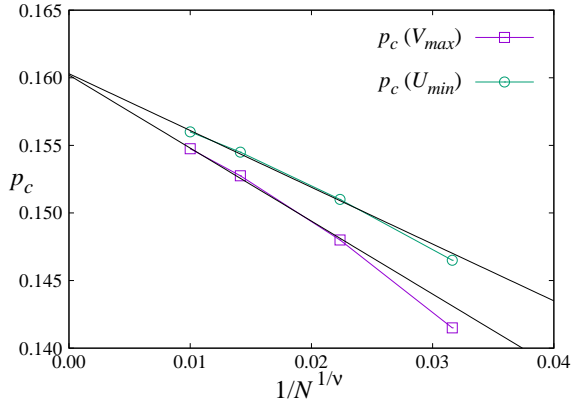}
\end{center}
\caption{Discontinuous transition in random regular networks, for $q=12$, $q_0=10$, without repetition. 
Values of $p$ for maximal susceptibility and minimal Binder cumulant 
as a function of $1/N^{1/\nu}$, with $\nu=2$. 
The critical point, $p_c=0.16025$, was obtained by extrapolation to $1/N\to 0$.
}
\label{fig:fss-discontinuous}
\end{figure}

For constructing the diagrams in Fig.~\ref{fig:diagrams} for random regular networks, the critical 
points were determined with accuracy of order $\pm 0.001$.
Finite-size scaling analysis was performed, as illustrated in 
Figs. \ref{fig:fss-continuous} and \ref{fig:fss-discontinuous}, 
for continuous and discontinuous transitions, respectively.

When the O-D transition is continuous (lilac-orange) the critical values and critical exponents were determined by defining the following quantities. The order parameter 
\begin{equation} \label{order3}
O = \left\langle \frac{1}{N}\left|\sum_{i=1}^{N} o_{i}\right|\right\rangle ~, 
\end{equation}
where $o_i=2x_i-1$, $\langle\, ...\, \rangle$ denotes a configurational average. 
It is sensitive to the unbalance between extreme opinions and 
plays the role of the ``magnetization per spin'' in magnetic systems. 
In addition, we also consider the fluctuations $V$ of the order parameter (or ``susceptibility'') 

\begin{equation} \label{chi}
V = N\,(\langle O^{2}\rangle - \langle O \rangle^{2}) 
\end{equation}
and the Binder cumulant $U$, defined as \cite{binder}
\begin{equation} \label{eq5}
U = 1 - \frac{\langle O^{4}\rangle}{3\,\langle O^{2}\rangle^{2}} \,.
\end{equation}

The critical value of $p$, the exponents $\beta$, $\gamma$ and $\nu$, 
were obtained by means of the usual scaling equations
\begin{eqnarray} \label{eq6}
O(N) & \sim & N^{-\beta/\nu}, \\ \label{eq7}
V(N) & \sim & N^{\gamma/\nu}, \\ \label{eq8}
U(N) & \sim & {\rm constant}, \\ \label{eq9}
p_{c}(N) - p_{c}& \sim & N^{-1/\nu} ~,
\end{eqnarray}
that are valid in the vicinity of the transition. 

The procedure is illustrated in Fig.~\ref{fig:fss-continuous}, for the case 
$k=q=12$, $q_0=7$, in random regular networks. 
We obtained mean-field values of the exponents $\beta=1/2$, $\gamma=1$, $\nu=2$.

The procedure for discontinuous transitions is illustrated in 
Fig.~\ref{fig:fss-discontinuous}~\cite{discontinuous1,discontinuous2}. 
We considered the same parameters as before. In particular the critical point 
was obtained by extrapolating the extreme values of $U$ and $V$ to 
the limit $N\to \infty$.



\begin{thebibliography}{40}
\bibitem{software} We used to this end the Wolfram Corporation Mathematica software.
\end{thebibliography}
\end{document}